\documentclass[11pt]{article}
\usepackage{arxiv}
\usepackage[utf8]{inputenc} 
\usepackage[T1]{fontenc}    
\usepackage{hyperref}       
\usepackage{url}            
\usepackage{booktabs}       
\usepackage{amsfonts}
\usepackage{graphicx}
\usepackage{amsmath}

\usepackage{bm}
\usepackage{dcolumn}
\usepackage{pdfpages}
\usepackage{verbatim}
\usepackage[export]{adjustbox}
\usepackage{chngpage}
\usepackage{floatrow}
\usepackage{longtable}
\usepackage{lscape}
\usepackage{amsmath}
\usepackage[tableposition=top]{caption}
\usepackage{subcaption}
\urldef{\viewsurl}\url{viewsforecasting.org}

\usepackage[natbib=true,backend=bibtex, style=authoryear, isbn=false, url=false, doi=false, maxbibnames=99, maxcitenames=2, firstinits=true,terseinits=true, eprint=false,dashed=false]{biblatex}

\renewbibmacro{in:}{%
  \ifentrytype{article}{}{\printtext{\bibstring{in}\intitlepunct}}}
\renewbibmacro*{volume+number+eid}{%
  \printfield{volume}%
  \printfield{number}%
  :
  \printfield{eid}}

\DeclareFieldFormat[article]{number}{\mkbibparens{#1}}
\DeclareFieldFormat[article]{pages}{#1}
\DeclareFieldFormat[article]{title}{#1}
\DeclareNameAlias{sortname}{family-given}
\DeclareDelimFormat{finalnamedelim}{\addspace\bibstring{and}\space}
\AtEveryBibitem{\clearfield{month}}
\AtEveryBibitem{\clearfield{day}}

\renewcommand{\shorttitle}{THE UNDERREPORTED DEATH TOLL OF WARS}

\title{The underreported death toll of wars:  a probabilistic reassessment from a structured expert elicitation}

\author{Paola Vesco \\
        Peace Research Institute Oslo \\
        \texttt{paoves@prio.org} \\
	\And
	David Randahl \\
	Dep. of Peace and Conflict Research \\
        Uppsala University\\
	\texttt{david.randahl@pcr.uu.se} \\
	 \And
	 Håvard Hegre \\
  Peace Research Institute Oslo \\
	Dep. of Peace and Conflict Research\\
	Uppsala University\\
	\texttt{hhegre@prio.org} \\
        \And
        Stina Högbladh \\
        Dep. of Peace and Conflict Research \\
        Uppsala University\\
	\texttt{stina.hogbladh@pcr.uu.se} \\
        \AND
        Mert Can Yilmaz \\
        Dep. of Peace and Conflict Research \\
        Uppsala University\\
	\texttt{mertcan.yilmaz@pcr.uu.se}
 }
 
\begin{document}

\maketitle

\begin{abstract}

Event datasets including those provided by Uppsala Conflict Data Program (UCDP) are based on reports from the media and international organizations, and are likely to suffer from reporting bias. Since the UCDP has strict inclusion criteria, they most likely under-estimate conflict-related deaths, but we do not know by how much. Here, we provide a generalizable, cross-national measure of uncertainty around UCDP reported fatalities that is more robust and realistic than UCDP's documented low and high estimates, and make available a dataset and R package accounting for the measurement uncertainty. We use a structured expert elicitation combined with statistical modelling to derive a distribution of plausible number of fatalities given the number of battle-related deaths and the type of violence documented by the UCDP. The results can help scholars understand the extent of bias affecting their empirical analyses of organized violence and contribute to improve the accuracy of conflict forecasting systems.

\end{abstract}


\noindent \textbf{Funding}: The research was funded by Riksbankens Jubileumsfond through the programme \textit{Societies at Risk}, Uppsala University, the European Research Council under  Horizon Europe (Grant agreement No. 101055176, ANTICIPATE), the Research Council of Norway (Grant agreement n. 334977, UFFAC), and the Center for Advanced Study (CAS) at The Norwegian Academy of Science and Letters.
\vspace{-3mm} \\

\noindent \textbf{Acknowledgements}:
Thanks to all the UCDP coders that have contributed their vast experience through workshop participation and questionnaire responses!
For more information on the UCDP and VIEWS and UCDP projects see \url{https://ucdp.uu.se} and \viewsurl.
\vspace{-3mm} \\

\newpage
\section{Introduction}

Empirical research on peace and conflict has increasingly relied on event datasets that provide information on armed conflict events and related deaths for multiple countries and over time. The majority of these datasets are based on automatic or manual coding of events from news and media sources, thus giving rise to potential reporting bias. However, common coding procedures lack a systematic approach to quantify the uncertainty about the reported values.

In general, there are two types of uncertainty and bias in conflict event datasets. One relates to the unknown and uncertain events that fail to be reported at all in the source material available; the other, to bias and uncertainty in the events  that are actually reported by credible news sources, government and NGO reports, etc. In this article, we will focus on the latter source of bias and uncertainty.

The Uppsala Conflict Data Program (UCDP) is one of the most acknowledged and widely used among such datasets, reporting organized violence globally throughout the period 1989--2023. The UCDP data are hand-collected systematically, based on a rigorous and immutable definition of armed conflict as a set of events where the use of armed force results in at least 25 battle-related deaths in one calendar year \citep{Davies2024JPR}. 
The adoption of a rigorous definition of armed conflict, coupled with the application of systematic and transparent coding rules, enable UCDP to provide data that are comparable across cases and countries and over time. UCDP relies on local, national, and international news sources, as well as on IGO, NGO and research centers' monitoring efforts. The UCDP has unrivalled experience and allocate considerable resources to assess reporting bias, using multiple  sources for each event wherever feasible.\footnote{For instance, using media and non-media sources may attenuate bias \citep[][]{Dietrich2020InternationalInteractions}, as different information sources portray specific aspects of and different statistical patterns in violence \citep{Davenport2002JCR}.}  Still, some bias is unavoidable. The magnitude of the reporting bias and uncertainty around the number of reported fatalities is hard to measure, and remains largely unknown. When in doubt, the UCDP is conservative in their estimates of deaths. Consequently, UCDP has a known tendency to be biased downward, but little is known about the amount of this bias, as the UCDP does not provide a simple, consistent estimate of their own uncertainty about the true number of deaths in the events they report.

In the present study, we aim to leverage the UCDP's own expert knowledge to construct a generic model that can generate probability distributions of plausible fatality levels for all possible events and related fatalities coded in the UCDP dataset. We use this model to estimate the probability distribution of the most plausible number of fatalities for any conflict event reported by UCDP. We utilize an expert elicitation approach, combined with statistical simulations, to provide a probability distribution around the best estimate of deaths recorded in the UCDP dataset. As the elicitation study gives us expert estimates of the uncertainty around reported fatalities by violence type and for varying contexts and countries, the resulting estimates of uncertainty can be generalized across countries and settings.

We find that the UCDP coders belief of the underlying distribution of actual fatalities given a UCDP record of fatalities is best approximated by a gumbel distribution with excess density located at the reported value itself. Given this distribution and the responses from the coders, our analysis suggests that the UCDP coders believe that fatalities are generally under-reported, with the relative rate of under-reporting decreasing non-linearly with the number of fatalities.\footnote{Relative rate of mis-reporting is calculated as the mean of the predicted distribution divided by the reported number of fatalities.} For low-fatality events, with a best estimate of 100 fatalities or less, the rate of under-reporting decreases from around 100\% for events with one fatality to around 30\% for events with 100 fatalities. For very high fatality events, in excess of 23,000 fatalities, the best fitted coder distributions indicate that the best fatality estimates are on average slightly over-reported although a non-null likelihood of under-reporting remains.

Our study contributes to the existing research on measurement uncertainty and media bias in conflict events by augmenting the UCDP `best' deaths estimates with distributions of plausible values of fatalities as assessed by the UCDP coders themselves, based on their knowledge of the coding procedures and news sources. Our approach yields two useful additions to the UCDP data: (1) a more general and useful estimate of uncertainty than the current `low' and `high' estimates -- which do not represent confidence bounds; and (2) a countering of UCDP’s tendency to publish conservative estimates of fatalities. 

Our simulation approach is easy and fast to replicate, and has high potential to be applied in other contexts. To facilitate the use of our uncertainty model, we have created the R package \texttt{uncertainUCDP} which can be used to obtain random draws of the plausible number of for any arbitrary UCDP event. The functions estimates the $\tilde{\theta}$ parameters of the conditional distribution given the reported number of fatalities and type of violence of an event based on the regression models for $\tilde{\theta}$ presented in the paper. This conditional distribution can then be used to obtain quantiles, densities, and random draws of the distribution, as well as other quantities such as the mean of the distribution. The user can also extract the $\tilde{\theta}$ parameters of the conditional distribution. We also make a set of draws from the package available as a downloadable dataset.

By improving our understanding of uncertainty around conflict deaths in UCDP data, this approach will open new research avenues on the implications of measurement error for the statistical analysis of conflict data. The uncertainty and bias documented here will certainly lead to biased coefficients. Moreover, the approach will improve the accuracy of conflict forecasting systems by correcting the conservative bias as well as integrating uncertainty around the central value.

\section{Uncertainty in conflict data}

We can distinguish two main types of uncertainty in conflict fatalities: first, there exists an inherent uncertainty around conflict events that are completely missing from any source of information. This is the so-called `fog of war', or the inherent uncertainty in conflict fatalities related to missing people that are not recorded as dead, or conflict events that are completely undocumented -- what we here refer to as `unknown and uncertain' events.

A second type of uncertainty concerns conflict events that are in fact documented, but for which the number of related fatalities is not completely certain. Sources of this latter type of uncertainty can arise from the coding process itself, for example from the application of UCDP's definition of organized violence, given the amount of information contained in the original source. This source of uncertainty may concern different definitional aspects, such as the count of fatalities, the actors involved in the conflict, the type of violence (one-sided, state-based, or non-state), and the geographic and temporal precision of the event. In some cases, UCDP has a good estimate of the number of deaths but cannot document the link to the conflict: for example, UCDP has information on the number of people killed in the Rwandan genocide owing to visual reporting of mass graves, but they are not necessarily able to link this information to the specific incidents or actors. Uncertainty around the number of fatalities in documented conflicts can also stem from the selection of information sources. Some sources may over- or under-estimate deaths, potentially leading to under- or over-reporting of deaths for the overall affected population or for a specific group. If the sources UCDP consults miss out on some events that would fall under their definition of conflicts, the number of fatalities will be underestimated. If the sources inflate the number of deaths, over-reporting will be more likely. This uncertainty is dependent on the amount of available information on a conflict, which in turn is affected by freedom of the press, access to communication technology, and political regime \citep{Drakos2006JCR, Croicu2017PoliticalResearchQuarterly}. In this study, we address the latter type of uncertainty, and we aim at estimating the uncertainty in fatalities related to documented conflict events. By contrast, our approach does not intend to answer the broader question of how many conflict events are completely undocumented and therefore missing from the UCDP dataset \citep{Price2014USADK}.

Many recent studies have investigated the data generation process (DGP) of conflict event data to understand the uncertainty in conflict fatalities \citep[see e.g.][]{Baum2015JournalofPeaceResearch, Weidmann2016AJPS, Croicu2017PoliticalResearchQuarterly, Dietrich2020InternationalInteractions}. There is, however, surprisingly little existing literature on the uncertainty of data \textit{given reports}.  Existing approaches mostly focus on the `unknown and uncertain' events, or conflict episodes that are completely unreported and do not even surface to the media. This strand of literature aims at uncovering the unreported bouts of violence by comparing extant event datasets coded from the news with `ground truth' data from non-media sources, including human right organizations \citep{Ball2018CHANCE}, truth commissions and international criminal courts \citep{HooverGreen2019DemRes, Tabeau2013CountingCivilianCasualties}, trade unions \citep{Guzman2012}, NGOs \citep{Sloboda2013CountingCivilianCasualties}, memory books \citep{Kruger2014}, and research centers \citep{Price2015Significance}. These approaches are internally valid and they have significantly advanced our knowledge of the `true' number of people killed by conflicts, by illuminating the reporting challenges posed by wars. 
Yet, considerably less is known about the magnitude of uncertainty in fatalities related to conflict events that are reported. Although these two types of uncertainty are inter-connected, they are also inherently different and thus require a distinct approach to be measured. Our study deals with the latter type of uncertainty, but for the sake of the reader, we summarize the literature on both types here.

A prevalent approach used by scholars to infer the true number of fatalities related to unknown and uncertain events involves comparing event data to a `ground truth´, i.e. a detailed and reliable records of deaths, mostly at the individual level, which may be obtained by matching different sources of information on single events such as from truth commission, development organizations, IGOs, research centers, or a combination thereof \citep{Ball2002, Ball2018CHANCE, HooverGreen2019DemRes}. An example of `ground truth' is the Kosovo Memory Book, an account of the death toll during the armed conflict in the Former Yugoslav Republic in 1998-2000 compiled by the Humanitarian Law Centre in Belgrade, which is considered to report nearly all deaths occurred during the conflict \citep{Kruger2014}. This approach has advanced our understanding of the magnitude of reporting bias and illuminated how many conflict episodes can go completely undocumented. A fundamental challenge of this approach, however, is that it requires an accurate and full representation of the conflict and the number of people killed, which we rarely have access to. Due to its time-consuming and labour-intensive nature, this approach generally focuses on individual conflict events, especially those that are terminated.

Matching the UCDP GED, the geo-referenced version of the UCDP database \citep{Sundberg:2015ged}, with data from US military’s Significant Activities (SIGACTS) to estimate the patterns of deaths in Afghanistan, \citet{Weidmann2015JCR} finds that only 53\% of the events overlap in the two datasets. In a later study of Afghanistan, \citet{Weidmann2016AJPS} show that only 28.5\% of the deadly insurgent-initiated events in 2008 are reported in international news. In a study of violence in Colombia between 1999 and 2008, \citet{Guzman2012} similarly estimate that up to 30\% of trade unionists killings were not reported. \citet{Sloboda2013CountingCivilianCasualties} compares estimates of deaths from the Iraqi Body Count against non-journalistic sources of information on violent deaths in Iraq. They find that journalistic sources tend to under-report incidents characterized by low violence intensity, but are more likely to report violent episodes involving civilians, compared to military sources which predominantly record combatant deaths \citep{Sloboda2013CountingCivilianCasualties}.

\citet{Restrepo2006JPR} compares cross-country, time-varying datasets with microdata on the Colombian conflict (CERAC), and find that cross-country sources largely under-reports conflict deaths. They also show that UCDP's under-reporting is largely driven by their definition of armed conflict, which excludes attacks against civilians and activities from illegal right-wing paramilitary groups. However, UCDP's more recent reporting of one-sided violence, which was not available at that time, would now cover at least part of these deaths.

\citet{Tabeau2013CountingCivilianCasualties} compares different estimates of the death toll in Bosnia-Herzegovina (1992–1995) and of the Khmer Rouge victims in Cambodia in the 1970s. They find that estimates of excess deaths range from about 25,000 to 329,000 during the Bosnia-Herzegovina war, and from 75,000 to 2.2 million in Cambodia. The striking difference across these estimates is likely due to lack of information (especially for Cambodia), but also to potential double-counting, poor source reliability, and over-reliance on a single source.

A study of the Syrian conflict by \citet{Price2015Significance} underscores how the completeness of deaths reporting varies across locations and periods, such that changes in the recorded data may reflect reporting dynamics more than the underlying conflict dynamics. For example, human rights reporting are affected by the presence of human rights groups being higher in areas under the control of one party to the conflict relative to the other party, leading to a more complete documentation of violence in some areas \citep{Price2015Significance}. In a study of the war in Ukraine, \citet{Radford2023PNAS} compare estimates of deaths and casualties from almost 5,000 reports. They find that Russia has suffered more deaths than Ukraine but under-reports their deaths by 30\%, and that both sides overestimate the losses suffered by their opponent.

The estimation of the uncertainty in fatalities related to known conflict events, by contrast, does not require the compilation of all possible sources of information around a conflict, but rather involves inferring the precision of a documented event, for example by relating the media coverage to country or local attributes to uncover parallel trends, or by exploiting information available from the coding process of the event datasets themselves to draw inference about events precision and uncertainty. These studies mainly focus on detecting potential drivers of uncertainties and biases in conflict fatalities, but they do not generally aim at estimating the magnitude of these uncertainties.

For example, \citet{Croicu2017PoliticalResearchQuarterly} leverage the precision scores assigned to each event in the UCDP GED dataset, which measures the level of detail of the recorded event, to study whether the ability of the press to access information can account for spatial variations in the precision of conflict data. They find that the coding precision about events decreases with the distance from internet access, suggesting that events that occur in areas where journalists have easier access to information are reported more accurately than events in the periphery.
\citet{Dietrich2020InternationalInteractions} exploit variation in the type of information source which is indicated during the coding process, and find that media coverage is associated with contextual factors such as international trade and access to communication technology, while the inclusion of non-media sources can reduce the extent of reporting bias by covering additional events.

\citet{Baum2015JournalofPeaceResearch} compares incident-level data on organized violence to textual information from international newspapers during the Libyan revolution in 2010-2011. They find that reporting bias is associated with regime type: media in non-democracies under-reports protests and nonviolent collective action by regime opponents, ignore government atrocities, and over-report those perpetrated by rebels, while the opposite patterns characterise democracies.
Similarly, \citet{Drakos2006JCR} analyses terrorist incidents for 1985-1999 and relates the probability of an attack being reported to the country's political regime and freedom of the press. The study finds that the magnitude of the reporting bias is influenced by regime type: due to the strong correlation between regime type and press freedom, highly democratic countries tend to over-report terrorism, while autocracies largely under-report it.

\citet{Ariyanto2008AJC} analyses information from local newspapers reporting on the Christian–Muslim conflict in Ambon, Indonesia, and uncovers the existence of a `naming bias', whereby both Christian and Muslim newspapers are more likely to blame religious outgroup members as perpetrators of violence than they are to attribute responsibility to their in-group. The prevalence of this bias is found to vary across the two religious groups: it is higher in the Muslim newspaper than in the Christian one.

In a study of the Guatemalan conflict, \citet{Davenport2002JCR} finds that different information providers portray distinct types and aspects of repression, depending on the goals of the observer, the characteristics of the event, and the overarching political context, such as the type of regime, the year in which the event occurred, and the number of human rights organizations active in the country at the time. In a study of media coverage of protest events, \citep{Herkenrath2011IJCS} finds that the extent of international coverage of protests depends largely on the country in which the event takes place.
\citet{Weidmann2015JCR} investigates the accuracy of reported UCDP GED events in Afghanistan and finds that the geographic precision of the events is related to the distance to the nearest major settlement. Analysing the UCDP GED dataset on Afghanistan for 2009, \citet{Otto2013CC} finds indication of both a significance bias towards the victims, such that the killing of a political figure is more likely to be reported than the killing of unknown citizens, and an omission bias arising from uncertainties in the application of the definition of one-sided violence. \citet{Gritten2012SJFR} studies 14 forest conflicts involving forest industry and relates their intensity and impacts to media coverage. The authors find that there is little international media coverage of the conflicts, and that the correlation between media coverage and conflict intensity or impact largely varies across contexts. \citet{Urlacher2009ISP} relates trends in civil war and journalist killings to news coverage, but fails to find a strong relationship, suggesting that news coverage is relatively robust to violence intensity.

These approaches have greatly advanced our knowledge of reporting bias and measurement uncertainty in conflict event datasets, with important implications for the empirical study of violence. However, they are mostly concerned with estimating the true, undocumented number of fatalities related to the `unknown and uncertain' events,  or to gain insight on the drivers of uncertainty in fatalities related to documented conflict events. To the best of our knowledge, a systematic effort to provide a measure of uncertainty in fatalities for events that are documented has yet to be undertaken.  

Here, we fill this gap by augmenting UCDP estimates of deaths with a probability distribution of plausible values of fatalities, for any fatality level recorded in the UCDP dataset. The resulting distributions provide a more realistic and robust confidence interval around UCDP best estimates than their documented high and low estimates -- which do not represent confidence bounds. We leverage on the expert knowledge of the UCDP team coupled with statistical simulations to obtain an estimate of uncertainty that is generalizable across information contexts, settings, and violence types. We detail our approach in the next section.

\section{Methods}

\subsection{UCDP coding of fatalities}

The Uppsala Conflict Data Program (UCDP) defines armed conflict as a violent event where the use of armed force leads to at least 25 battle-related deaths at the country level in one calendar year \citep{Gleditsch2002JPR, Davies2024JPR}.
UCDP consistently reports on three types of organized violence: state-based conflict, where the use of of armed force involves at least one governmental actor; non-state conflict between two organized groups, none of which is the government of a state, and one-sided violence, where a government actor or a non-state group deliberately targets civilians.

For each event and each type of violence, UCDP reports a `best estimate' of the number of people killed, which represents the most reliable count of deaths associated to that particular event \citep{Hogbladh2023codebook, Sundberg2013JPR}. We will refer to this as the `reported value' for the event, denoted $\tilde{y}_i$, where $i$ is an index for the specific event. In addition to the reported value, each event record includes a `low' and `high' estimates of fatalities. As a rule of thumb, the best estimate reflects the most reliable and documented estimate of the number of fatalities. However, the UCDP is admittedly conservative in their reporting \citep[e.g.][]{UCDP2023Syria}, and it is plausible that the true number of fatalities is closer to their high estimate than the best estimate, depending on the context of an event and the country in which it occurs.

It is important to note that the UCDP low and high estimates do not represent a confidence interval around the best estimate: they are the result of a systematic coding process, and reflect the uncertainty and reliability of available sources, but they are not an estimate of measurement error.\footnote{Appendix Table \ref{tab:UCDP_coding} summarises how the coding of best, low, and high estimates in the UCDP dataset are used to identify different types of uncertainty around the event, the type of violence, the conflict actor, or the reliability of the information source. For example, cases for which both the low and best estimates are coded as 0 but the high estimate is non-zero, intentionally identify uncertainty about the type of violence or actors. In around half of the recorded observations, the low estimate is coded as equal to the best estimate.}

UCDP distinguishes two types of conflict events by `clarity' \citep[precision, cf.][p. 22]{Hogbladh2023codebook}: `ordinary' individual events with high clarity (1), and `summary' events with lower clarity (2). Events coded with clarity 1 are those for which the source is sufficiently detailed to identify individual incidents in a single location and date. Summary events are events for which the source is aggregated to an extent that makes it impossible for the coder to disentangle: these are aggregations of incidents over a longer period than a single, clearly defined date and a specified point in space \citep{Croicu2013ged}. To avoid double-counting, the fatalities reported in `summary' events are compared with the fatalities from individual incidents occurred in the same location and time span, and the fatalities from individual events are subtracted from the fatalities reported in summary events. This protocol, combined with the other coding rules adopted by UCDP, implies that in few cases the high estimate of fatalities is lower than the best fatality estimate.\footnote{This happens in around 1\% of UCDP recorded events.}

Although the low, best, and high estimates that UCDP codes and publishes contain highly informative and valuable information which is carefully documented \citep[see][\url{https://ucdp.uu.se/downloads/ged/ged241.pdf }]{Hogbladh2023codebook}, this information is conveyed in a form that is specific to distinct sub-cases, and is not a practically useful or accessible representation of uncertainty. The UCDP coding protocol also contains a number of other practices  based on an understanding of uncertainty which is documented, yet not easily accessible by users. For instance, a news report stipulating `many fatalities' will be coded as 3 fatalities by the UCDP according to their practice. All UCDP coders know in this case that a coding of `3 fatalities' in many cases means `3 or more' rather than `3'. A coding of `4 fatalities', on the other hand, is more precise due to this coding convention. These rules are encompassed in the so-called `vague number translator' reported in the Supplementary Material.

The expert elicitation we describe below taps into the UCDP knowledge about all of these documented but internal coding practices, in addition to their accumulated experience and understanding about the uncertainty in the reports themselves. We seek to present the UCDP coders' uncertainty in the form of probability distributions over all possible fatality counts, given the published report. This format has the advantage of being uniform across all cases, but ignores the detailed information about the sources of the uncertainty in the original UCDP coding.

\subsection{Expert elicitation}

We use an expert elicitation approach to extract UCDP coders' probabilistic assessments of plausible fatalities -- the `coder distributions' given the reported value and the type of violence recorded in the UCDP dataset. These assessments are used to quantify the uncertainty around each recorded `best estimate' of fatality in the UCDP dataset. The expert elicitation coupled with statistical modelling enable us to create a distribution of plausible values for the unreported number of fatalities, given the reported ones. The elicitation process is described as follows.

First, we conducted a workshop with 15 UCDP coders to discuss the main concepts, delimit the scope and ambition of the survey and pilot it. The UCDP coders undergo extensive training on the coding rules and procedures, and participate in weekly team meetings to discuss difficult cases, which increases inter-coder reliability. In our sample -- encompassing the majority of the UCDP team -- coders have 2-5 years of experience, although considerable variations exist, with 3 out of 15 coders having more than 10 years of experience, and 2 coders having less than 1 year of experience.  

The main objective of the workshop was to gain knowledge about UCDP's coding procedures, understand what they believe could be known about the `true', i.e. the most plausible number of fatalities, given that a best estimate of the number of fatalities is coded in the main dataset, and shed light on their understanding of the main factors affecting the availability and reliability of information sources throughout different contexts. In the first part of the workshop, we built a common understanding of the concept of uncertainty and how it varies across countries, conflict events and information sources, and primed UCDP coders with a non-technical understanding of probability distributions. Next, we piloted the questionnaire, explained how to fill it in through the platform, and received their feedback on how to clarify and improve it. In the second part of the workshop, we discussed how UCDP codes the low, best, and high estimates of conflict fatalities for any given event, based on the type of conflict, the actors involved, the available information sources, and other relevant characteristics of the event and the country in which it takes place. Lastly, we had an open guided discussion on the most uncertain conflict contexts, such as Ethiopia, where information on conflict events and related fatalities is lacking, biased, or not reliable.

The second stage involved developing the questionnaire design based on insights from the workshop, to be implemented in Qualtrics. The questionnaire was designed to elicit intervals rather than precise event probabilities.  The questionnaire covered all types of conflict events coded by UCDP, with a special focus on state-based violence -- the most prevalent type of conflict. The first part included country-level questions. We first asked the respondent to list all the countries they have coded for, including those that they are no longer coding but they have coded in the past. Next, we asked them to indicate which of the coded countries has the `best' information coverage and reliability, and which country has the `worst'. The best country represents a good information context -- i.e. a context where information on the conflict event is abundant, diverse and relatively reliable --, the worst, a bad one. For each country that they have coded for, the respondent was asked what they believe to be the most likely `true' level of fatality for the 2020--2022 period, including also those fatalities that are not reported by UCDP.

Next, we asked a set of calibration questions to understand which factors shape the availability and reliability of information, and to what extent. These factors, identified by the UCDP team during the workshop, include the type of violence, the type of information provider (media and non media sources, local and international news, NGO/IGO reports), the type of actor releasing the information (rebel groups, government), the political regime, the timing of the event and history of conflict (more or less recent conflicts), geographic characteristics (mountainous terrain, rural or urban area), and subjective characteristics of the coder such as their experience in coding. Additional calibration questions asked the respondents to compare the availability and reliability of information of a specific country against the region of relevance (e.g. Africa) as well as the world. To capture their beliefs on time variation, we also asked whether they believe that new or better information for each country would become available in the future and why.

\begin{figure}[htp]
    \centering
    \includegraphics[width=0.9\textwidth]{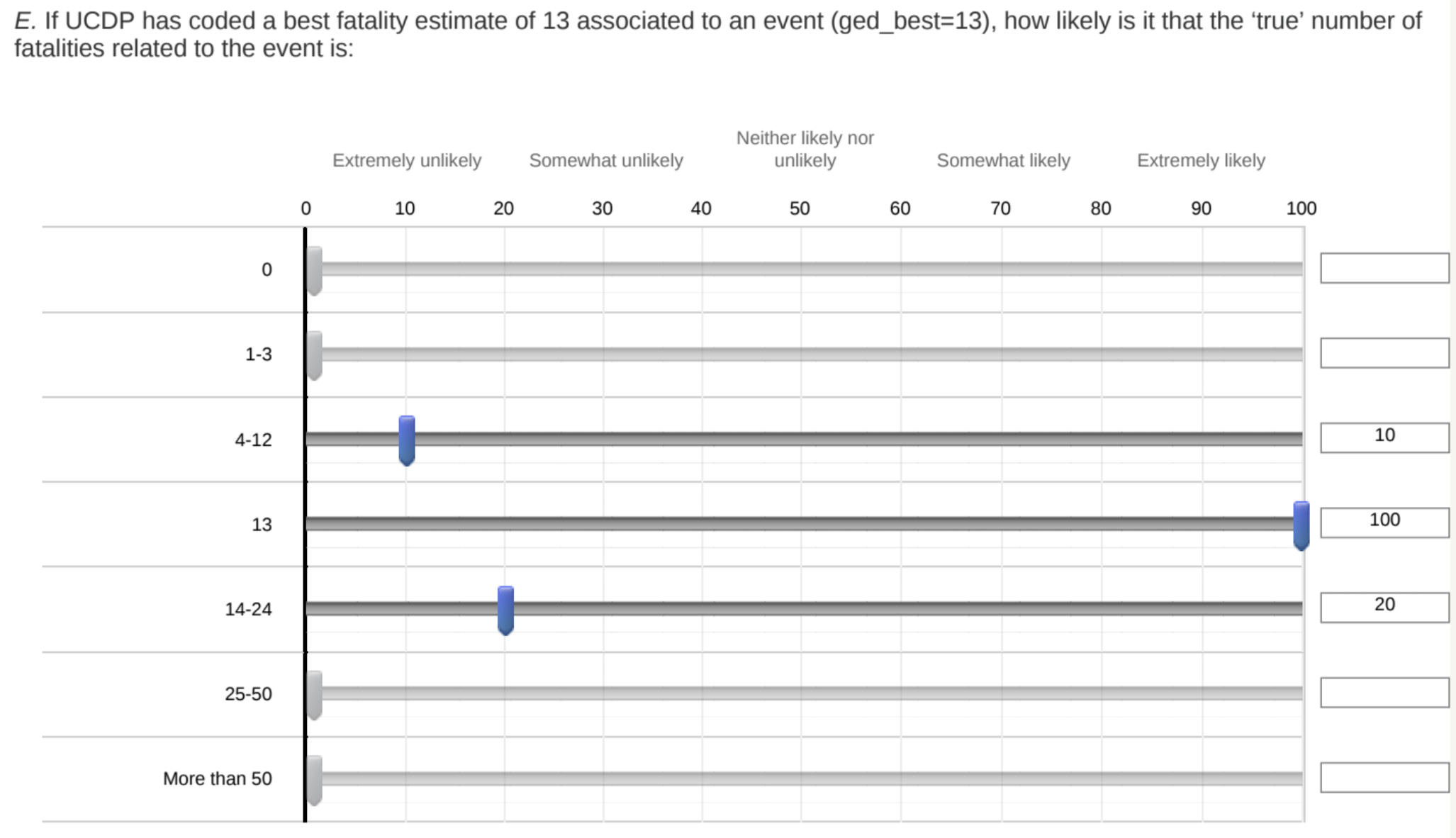}
    \caption{Example question and response options in the survey administered to the UCDP team, and resulting response for $\tilde{y}_i=13$}
    \label{fig:example_survey}
\end{figure}
The second part of the questionnaire included event-level questions, across both bad and good information contexts, where we retrieved what we will call `coder distributions' -- a representation of each coder's subjective probability density function (PDF) over the true number of fatalities for an event given how the UCDP coded it. We iterated the event-level questions over the three types of violence coded by UCDP. For a selected number of fatalities reported in the UCDP dataset -- best estimate equal to 0, 1, 2, 3, 13, 20, 24, 40, 47, 100, 101, 200, 201, 1000, 1001, 2000, 2001, 10,000, 100,000 -- we asked the respondents how likely it is that the `true' number of fatalities related to the event lies in a specified interval. The levels of fatalities for which we asked the questions were highlighted as significant by the UCDP team during the workshop and preliminary discussions, as they are relevant for their coding processes.
An example survey question and resulting coder distribution is presented in Figure \ref{fig:example_survey}, for an event recorded by the UCDP to involve $\tilde{y}=13$ fatalities.

The response intervals in the survey questions were selected according to the number of reported fatalities. For each interval, the response options were: Extremely likely (90–100\%), Somewhat likely (60-90\%), Neither likely nor unlikely (40-60\%), Somewhat unlikely (10–40\%), Extremely unlikely (0–10\%). To respond, the participants could slide the bar and drag it up to the appropriate response option. We clarified to the UCDP team that the response options did not have to sum to 100, as we would normalize their responses. The coders' responses thus represent binned probabilities of the plausible values of fatalities, given the specific fatality levels that were surveyed, and for different information contexts and types of violence. A sample coder distribution obtained from the survey is provided in Table \ref{tab:sample_response}.

\begin{table}[htp]
    \centering
    \begin{tabular}{l|r|r}
    Bin & Fatality range    & Bin probability \\
    \hline
    1   &  0                & 0.01 \\
    2   &  1--3             & 0.02 \\
    3   & 4--12             & 0.12 \\
    4   & 13                & 0.64 \\
    5   & 14--24            & 0.19 \\
    6   & 25--50            & 0.015 \\
    7   & $>$ 50            & 0.01 \\
    \hline
    \end{tabular}
    \caption{Sample coder distribution from individual coder $j$ for the example in Figure \ref{fig:example_survey}: UCDP reports $\tilde{y_i}=13$ fatalities, high-information context, state-based violence}
    \label{tab:sample_response}
\end{table}

The questionnaire was distributed to the entire UCDP team through a Qualtrics link and could be filled out from any laptop or smartphone. We instructed respondents to complete the questionnaire alone, without consulting their colleagues, as variation across their responses was valuable to us. We collected responses from 13 UCDP coders during the period May--June 2023.

The final stage involves a statistical analysis of the results to derive probability distributions of the plausible values of fatalities, for any level of reported fatality in the UCDP dataset, and across different information contexts. The next section describes the methods employed to obtain a distribution of the likely underlying fatalities for all UCDP violent events, as well as for aggregations of events, which we can use to produce a representation of the uncertainty around each UCDP reported level of fatalities, and assess the rate of over/under reporting.

\subsection{Statistical modelling}

As the survey asked a finite number of questions, we could collect the coders' probabilistic assessments only for a limited number of reported fatality values. However, we want to extrapolate this information for all possible values, beyond those surveyed, to extract a continuous probability distribution of plausible values for each UCDP best estimate. The analysis illustrated in the next sub-section enables expanding the probabilistic assessments from the levels of fatalities for which we asked a question in the survey to all integers equal to or greater than zero.

\subsubsection{Estimation strategy: a model of $y_{i}$}

The survey provides us with each UCDP coder's coder distribution for a sample of  values recorded as the UCDP best estimate (the reported value). A core aim of the analysis is to construct a generic model that can generate probability distributions of plausible fatality levels for all possible events and related fatalities coded by UCDP. If a UCDP coder has coded $\tilde{y}_i = 37$ fatalities in state-based conflict in a good-information context, we want to use the model to estimate the probability distribution of the most plausible number of fatalities for this event, even though this particular reported value was not included in the survey.

We can think of the distribution of plausible values for the true number of  fatalities of an event $i$ as a PDF with parameters $\bm{\theta}_i$, where $\bm{\theta}_i$ in turn is a function of the reported value $\tilde{y}_{i}$ and some contextual variables (e.g. information context), $\bm{z}_{i}$. The quantity of interest in this study can be then defined as:
 \begin{center}

\begin{equation}
    y_i \sim F(\bm{\theta}_i|\tilde{y}_{i}, \bm{z}_{i})
\label{eq:y}
\end{equation}

\end{center}
with:

\begin{center}

                           \begin{equation}
    \theta_{i,k} = f(\bm{\beta_k},\tilde{y}_{i},\bm{z}_{i},\epsilon)
\label{eq:theta}
\end{equation}
\end{center}
where $i$ indexes a record of a UCDP violent event; $y_{i}$ is the PDF for the plausible value of the true (unobserved) number of fatalities for $i$,  and $\theta_{k,i}$ is the $k$:th parameter of the PDF of violent event \textit{i} as a function of  $\tilde{y}_{i}$, the number of reported fatalities in UCDP for event $i$, and $\bm{z}_{i}$,  a set of contextual factors affecting uncertainty around $y_{i}$, e.g. factors related to the coding process or the general information context.\footnote{$Y_i$ and $\tilde{y}_i$ are limited to the set of non-negative integers smaller than the total population of the Earth.} For example, $\bm{z}_{ij}$ includes a set of dummy variables for reported counts that the UCDP coding processes treat as special values;\footnote{Specifically, we include a dummy for the following special values: 2, 3, 13, 30, 24, 40, 101, 200, 1001, 2000 which are defined in the UCDP's coding rules. These special values are included in the UCDP `vague number translator', a document that dictates the conversion between fuzzy textual terms from information sources to finite numbers in the coding of fatalities. For example, if a news article reports that `many' people have been killed, the vague number translator indicates that 3 fatalities should be reported for that event.} and $\epsilon$ is an error term. 

Our objective is to find the parametric distribution with parameters $\bm{\theta}_i$ which best mimics the binned probabilities assigned by the coders to the events in the survey -- which we refer to as the `coder distributions'. As we want to extrapolate a probability distribution of plausible values for all fatality levels, beyond those surveyed, our model needs to be able to generalize from the coders' distribution and predict $\bm{\hat\theta}_i$ for any level of fatality reported by UCDP, given some contextual factors $\tilde{y}_{i}, \bm{z}_i$. 

We achieve this through a two-stage process. In the first stage, we select a wide range of distributions and mixture distributions, and for each distribution, coder $j$, and event $i$, we obtain the `optimal' set of parameters $\bm{\tilde\theta}_{i,k}$, i.e. the parameters of the distribution which most closely approximate the coders' distribution. In the second stage, we estimate a new set of parameters $\bm{\hat\theta}_i$ for a distribution that generalizes from the $\bm{\tilde\theta}_{i,k}$ for each coder and each sample event.

As our goal is to translate the coders' binned distribution into a continuous distribution, we only consider continuous and mixture distributions, and not those which could mimic only the binned coders' distribution directly, e.g. the binomial distribution. We test the normal, lognormal, gumbel, negative binomial (all defined by two parameters), and poisson (one parameter) distributions, as well as reported-value inflated mixture versions of these distributions. The reported-value inflated mixture distribution is a weighted combination of a standard parametric distribution and a point distribution which has its entire density assigned to the reported value. The reported-value mixture distributions thus have an additional parameter, $w$, which controls the mixture weight between the parametric and point distributions. These distributions are introduced to account for the coders general belief that the UCDP recorded estimates are approximately correct, which is reflected in a much higher density on the exact reported value in the coders' distribution than would be expected by any standard parametric distribution. In addition, we also test a shifted version of all distributions where we shift the distribution to be centered at zero. Shifted distributions are characterized by an additional shift parameter, $s$ by which the distribution is shifted, for instance to allow for negative values.

In the first stage, we obtain the optimal $\bm{\tilde\theta}_{i,k}$ for each distribution, coder $j$, and event $i$ by running a genetic algorithm which spans the reasonable parameter space for each parameter of the distribution. For each set of suggested $\bm{\theta}$ we then compare the resulting parametric distribution with the binned coders' distribution \footnote{We bin the parametric distribution using the same bins as in the coders' distribution} and evaluate the fitness by taking the sum of the absolute differences between the coders' distribution and the binned parametric distribution. This difference, which we call Binned Absolute Difference (BAD), ranges from 0 to 2, where 0 indicates identical distributions and 2 designate completely separate distributions. In each generation of the algorithm, the fitness of the set of parameters is evaluated using the BAD score. To ensure that we obtain the optimal $\bm{\tilde\theta}_{i,k}$ we run the algorithm for 100 generations.\footnote{A genetic algorithm was chosen over numerical optimization techniques due to the complexities of obtaining correct derivatives for the mixture distributions and the BAD score. The algorithm seems to converge in the minimum BAD score well before reaching 100 generations so 100 generations is likely to produce (near) optimal $\bm{\tilde\theta}_{i,k}$}

Once we have obtained the optimal $\bm{\tilde{\theta}_{i,k}}$ for all combinations of distributions, coders, and events, we turn to the second stage of the process in which we model the optimal $\bm{\hat{\theta}_{i,k}}$ as a function of our contextual factors, $\tilde{y}_{i}, \bm{z}_i$. This stage allows us to extract the predicted parameters for the distribution for any event reported by UCDP, $\bm{\tilde{y}}_i$, conditional only on these contextual factors. In practice, we model these parameters as linear or logistic regression models, using different combinations of contextual factors as the independent variables and the  $\bm{\tilde{\theta}}_{i,k}$ optimized in the first stage as the dependent variable. For each distribution and combination of contextual factors, we then evaluate the out-of-sample performance of the model by running a leave-one-coder-out cross validation procedure; for each event in the survey we predict $\bm{\hat{\theta}}_i$ and calculate the BAD score,dropping one coder from the data at  each iteration. To evaluate the final performance of the distributions and combinations of contextual factors, we then take the median of the obtained BAD scores. \footnote{The median is chosen over the mean because of a few very low and high BAD values for certain distributions. However, the results remain nearly the same when using the mean instead.}

The process is summarized in Figure \ref{fig:process} below.

\begin{figure}[H]
     \includegraphics[width = 0.99\linewidth]{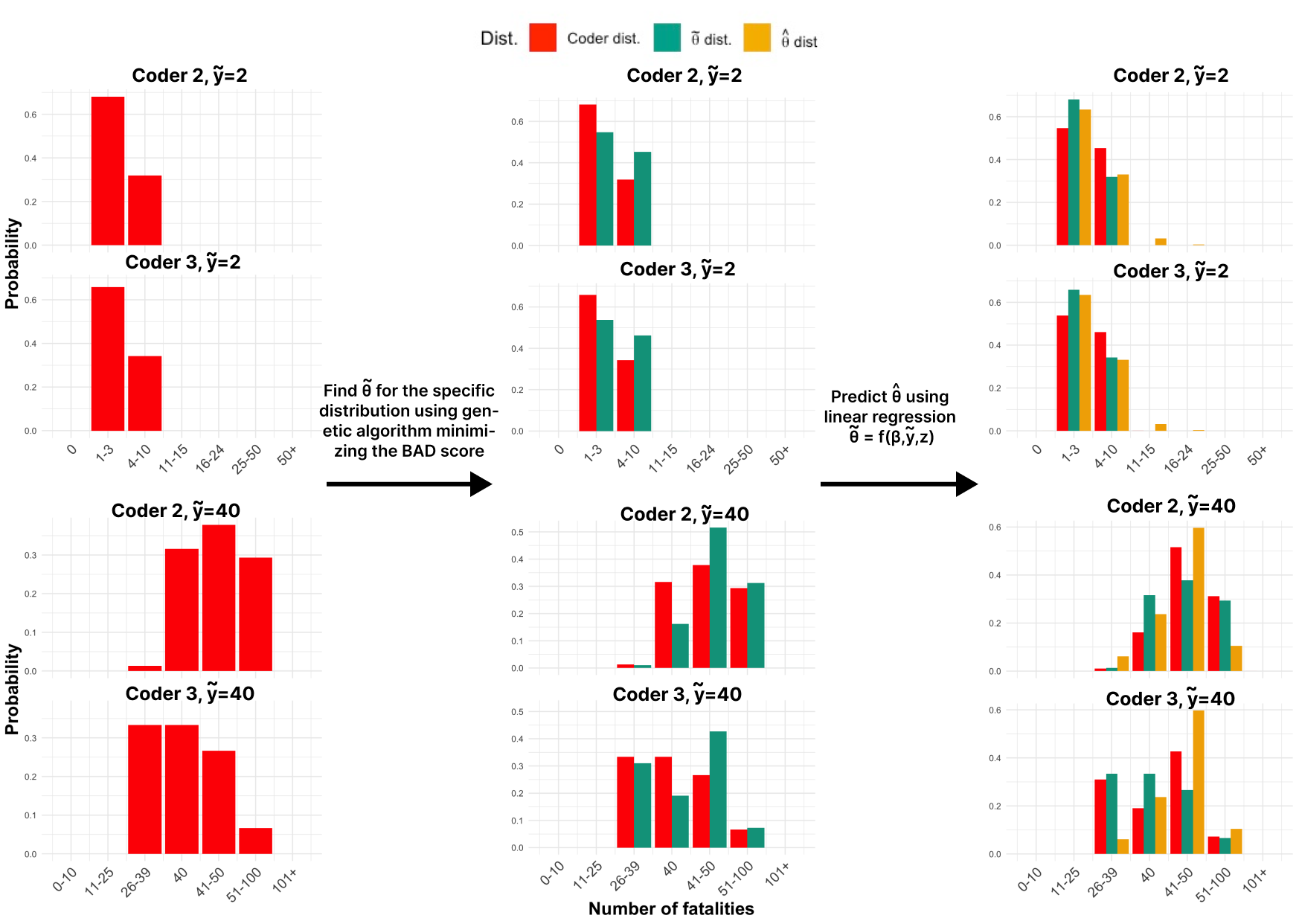}
     \caption{The two-step procedure for evaluating any given distribution, presented for two example coders and two fatality levels and for the point-value inflated lognormal distribution}\label{fig:process}.
  \end{figure}

\section{Results}
The mean and median BAD scores for the \textit{best performing} specification per distribution and type of violence are shown in Tables A2-A5 in the Appendix. The results show relatively small differences between the distributions, but suggest that the reported-value inflated gumbel mixture distribution is the distribution which produces the lowest aggregate median BAD scores. This holds across all types of violence. It is noteworthy that the reported value inflated mixture distribution outperform their non-mixture counterparts across all distributions and types of violence. As the coders generally believe that the UCDP reported value is close to the true value, it is unsurprising that the reported-value inflated mixtures overall outperform their non-mixed counterparts. Since the reported value inflated gumbel mixture distribution is the best performing of the distribution across the types of violence we will focus on this distribution for the remainder of the paper.

The reported value-inflated gumbel mixture distribution has three parameters, the location parameter $\theta_1 = \mu_g$, the scale parameter $\theta_2 = \beta_g$ and the mixture weight parameter $\theta_3 = w$. In the second stage, where we predict $\hat{\bm{\theta}}$ from $\tilde{\bm{\theta}}$, we test a number of different combinations of covariates for the regression models to predict the three parameters. The mean and median BAD scores for the different types of violence and for different combinations of covariates can be seen in Tables A6-A9 in the Appendix. Focusing on state-based violence, in Table A6, we see that the best performing combination of covariates uses the natural logarithm of the reported value, $log1p(\tilde{y})$ as well as the set of dummy variables for special values, $D$, for all three parameters and the set of contextual variables, $z$, for the weight parameter. However, differences in the mean and median scores across different specifications of covariates are generally small. As we strive for a maximally generalizable model of the uncertainty in the UCDP data which can be applied to any arbitrary UCDP event, we choose the model specifications using $log1p(\tilde{y}), D$ as covariates and excluding $z$. This choice saves us from having to estimate $z$ for any individual event, while still ensuring high accuracy. Similarly, we choose to use $log1p(\tilde{y}), D$ as covariates for all three parameters of the distribution across all levels of violence, because different combinations of $log1p(\tilde{y})$ and $D$ feature among the best performing specifications across the different types of violence.  We therefore err on the cautious side and include both $log1p(\tilde{y})$ and $D$ for all parameters across all types of violence. 

The output for the regression models used to predict $\hat{\bm{\theta}}$ for the gumbel mixture distribution with this specification of covariates can be seen in Table \ref{tab:regression_table}.

\begin{table}[htp] \centering 
 
\begin{small}

\begin{tabular}{@{\extracolsep{5pt}}lccc} 
\\[-1.8ex]\hline 
\hline \\[-1.8ex] 
Model & \textit{linear} & \textit{linear} & \textit{logit} \\ 
$\theta$ & $log(\mu_{g}+1)$ & $log(\beta_{g})$ & $w$\\ 
\hline \\[-1.8ex] 
 $log1p(\tilde{y})$ & 0.933$^{***}$ & 0.831$^{***}$ & $-$0.051 \\ 
 $D_{2}$ & 0.117 & 0.181 & 0.494 \\ 
 $D_{3}$ & $-$0.233$^{*}$ & $-$0.259 & $-$0.232 \\ 
 $D_{13}$ & $-$0.402$^{***}$ & $-$0.510$^{*}$ & 0.340 \\ 
 $D_{20}$ & $-$0.377$^{***}$ & $-$1.345$^{***}$ & $-$0.114 \\ 
 $D_{24}$ & $-$0.469$^{***}$ & $-$1.587$^{***}$ & $-$0.659 \\ 
 $D_{40}$ & $-$0.280$^{**}$ & $-$1.283$^{***}$ & $-$0.109 \\ 
 $D_{101}$ & $-$0.213$^{*}$ & $-$1.009$^{***}$ & $-$0.065 \\ 
 $D_{200}$ & $-$0.209$^{*}$ & $-$0.636$^{**}$ & $-$0.025 \\ 
 $D_{1001}$ & $-$0.003 & $-$0.207 & $-$0.012 \\ 
 $D_{2000}$ & $-$0.066 & 0.298 & $-$0.045 \\ 
 Constant & 0.592$^{***}$ & $-$0.307$^{*}$ & $-$0.095 \\ 
\hline \\[-1.8ex] 
Observations & 448 & 448 & 448 \\ 
R$^{2}$ & 0.967 & 0.813 &  \\ 
\hline 
\hline \\[-1.8ex] 
\textit{Note:}  & \multicolumn{3}{r}{$^{*}$p$<$0.1; $^{**}$p$<$0.05; $^{***}$p$<$0.01} \\ 
\end{tabular} 
\end{small}
\caption{Regression coefficients for the parameters of the reported value-inflated gumbel mixture distribution. Columns 2--4 report the predicted values for the three $\theta_k$ parameters for this distribution: (1) $\mu_g$,the gumbel location parameter; (2) $\beta_g$, the gumbel scale parameter; and $w$, the proportion of the probability mass that is drawn from the point distribution at the reported value.}
\label{tab:regression_table}
\end{table}

The results from the regression analysis shows, as expected, that there is a strong linear relationship between the logarithm of the number of reported fatalities and the location parameter of the gumbel distribution. Crucially, the regression results implies that the UCDP fatalities are on average under-reported except for extreme levels of violence with $\tilde{y}>23,000$. The rate of under-reporting, defined as the mean of the parametric distribution divided by the reported value, is generally decreasing with the number of fatalities.\footnote{Some of the special numbers dummies break this pattern.} For reported fatality levels under 100, the estimated rate of under-reporting ranges from over over 100\% -- for events with 1 fatality -- to around 30\%  --for events with 100 fatalities. The results also show that the dummies for the vague numbers in the UCDP generally have negative effects on the two first parameters of the distribution, indicating that the coders believe that these \textit{vague numbers} are less prone to be under-reported than other numbers, except for the dummy for $\tilde{y}=2$. As an example, coders know that `dozens' of reported fatalities (corresponding to the dummy for $\tilde{y}=24$), especially in bad information contexts, might reflect a tendency of media sources to over-report fatalities. The coders thus associate a non-negligible probability that the reported fatalities in the UCDP dataset are higher than the actual values for such special cases.


The relationship between the overall mean of the reported-value inflated gumbel mixture distribution and the reported number of fatalities can be seen in Figure \ref{fig:mu} below. This figure shows that there is nearly a 1-to-1 increase between the predicted mean of the parametric distribution and the reported value. The Figure also shows a  relative higher rate of under-reporting for lower levels of $\tilde{y}$. The fitted line displays some peculiar dips in the mean corresponding to the special values defined by the UCDP vague number translator, in line with the regression results presented above. The dotted lines below represent the 97.5th and 2.5th percentiles of the distribution, and show that (on the logarithmic scale) the estimated uncertainty of fatalities for a given event decreases with the number of reported fatalities.

 \begin{figure}[htp]
     \includegraphics[width = 0.99\linewidth]{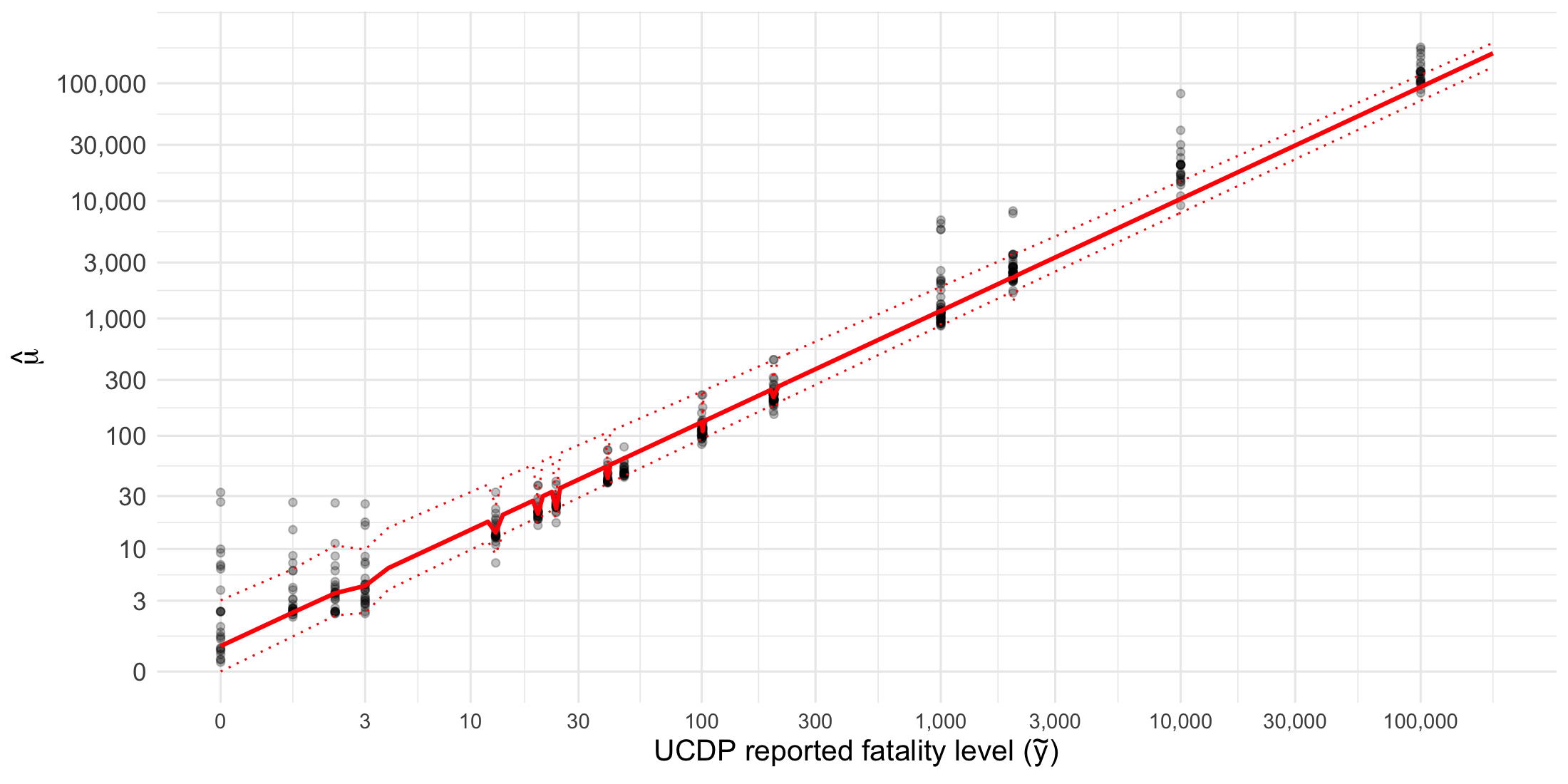}
     \caption{The line shows the predicted $\mu$, denoting the overall mean of the plausible fatality level ($y$ axis) for any given level of fatality reported by UCDP ($x$ axis), according to our preferred model specification. The dots represent the predicted means for each sample of $\tilde{y}_i$ for which we have coder distributions. The dotted lines represent the 97.5th and 2.5th percentiles of the resulting distribution.}
     \label{fig:mu}
  \end{figure}

We can also use the fitted regression models to produce densities of plausible actual values of fatalities given a specific reported value and type of violence, and for any given event in the UCDP dataset. Figure \ref{fig:densities} below shows the densities of the plausible number of fatalities for eight different $\tilde{y}$.

 \begin{figure}[htbp]
     \includegraphics[width = 0.99\linewidth]{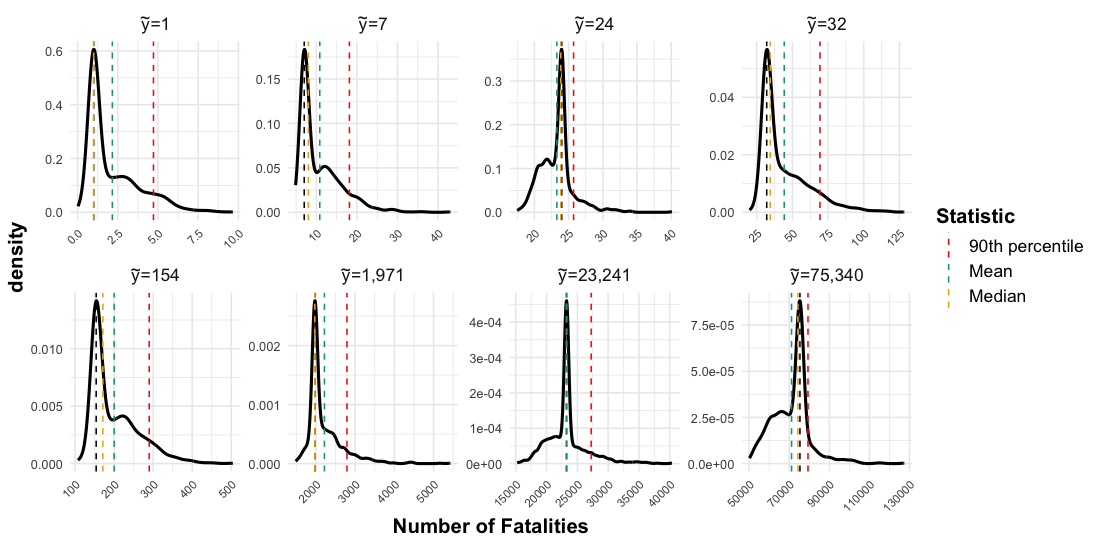}
     \caption{The facets each show a density for a specific $\tilde{y}$, the black dashed line correspond to $\tilde{y}$ and the colored dashed lines correspond to the median, mean, and 90th percentile of the distribution.}
     \label{fig:densities}
  \end{figure}

Figure \ref{fig:densities} highlights that the relative rate of under-reporting is higher for lower levels of violence, while for around 23,000 reported fatalities the mean and median of the distribution is the same as the reported value. For even higher number of fatalities , the mean and median of the predicted distribution is lower than $\tilde{y}$, although there remains a non-negligible risk of substantial under-reporting, with a 10\% likelihood that the number of fatalities is under-reported by at least 4,000.\footnote{75,340 is the $\tilde{y}$ of the event with the highest number of fatalities in Ethiopia on average in 2020--2022.} Figure \ref{fig:densities}  also shows a non-negligible risk of over-reporting for the special number $\tilde{y}=24$, corresponding to the UCDP coding for `dozens' of fatalities. This is particularly evident when comparing the distribution for this special value to the distribution of 32 fatalities.

\subsection{Distributions of fatalities across aggregated events}
Once we have identified our preferred parametric distribution and the best fitting regression models to predict the parameters of this distribution, we can use this information to predict a distribution of plausible fatalities related to any given UCDP event, given the number of fatalities $\tilde{y}$ reported by UCDP, and for a specific type of violence. To illustrate what information we can gain from this approach, we simulate plausible distributions of the total number of fatalities in Ethiopia and Syria in the years 2020--2022. Ethiopia and Syria are selected as two cases which represent a good (Syria) and bad (Ethiopia) information contexts in the selected time period. 

Specifically, we employ our estimation strategy to predict the distribution of plausible fatality level for all state-base conflict events reported by UCDP in each country in 2020-2022 Next, we make a random draw from the predicted distribution for each event and sum the total number of fatalities across all events. We repeat the procedure 1,000 times to obtain a distribution of the total plausible number of fatalities for each case. 

Figure \ref{fig:syr-eth} below shows the predicted distribution of the total number of fatalities in 2020--2022 for the two cases. The black dashed line represents the sum of fatalities in the UCDP, while the dashed orange and green lines represents the mean and median of the predicted distribution and the dashed red line shows its 90th percentile. 

\begin{figure}

    \includegraphics[trim={0, 0, 2cm, 0}, clip, width = 0.99\textwidth]{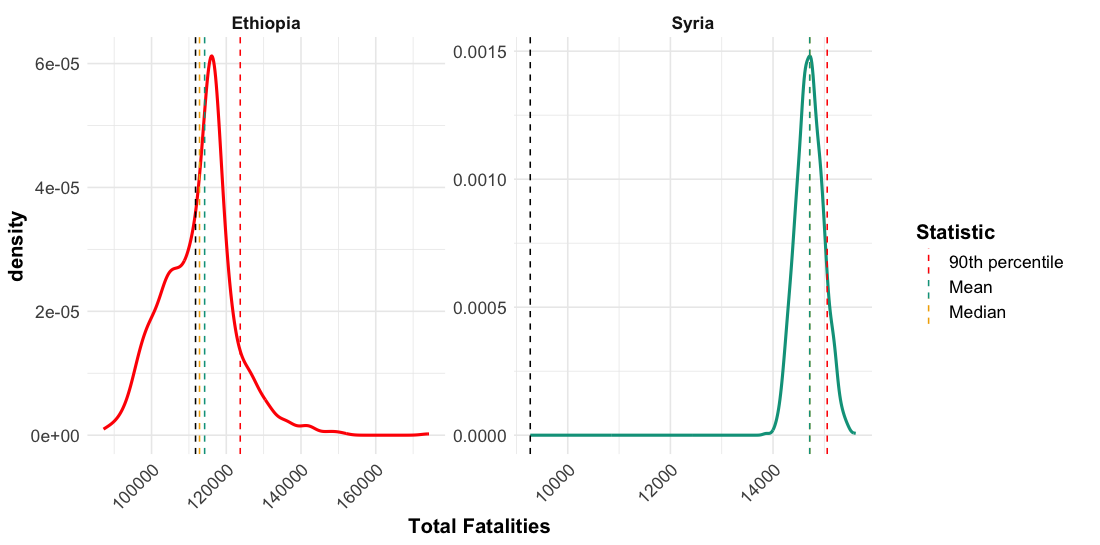}
    \caption{Distribution of plausible fatalities values for Syria (left) and Ethiopia (right) in 2020--2022, aggregated across all violent events. Black dashed line shows sum of the UCDP's `best' estimates, colored dashed lines the median, mean, and 90th percentile of the estimated distribution.}
    \label{fig:syr-eth}

\end{figure}

The results presented in Figure \ref{fig:syr-eth} show that the relative rate of under-reporting is much higher for Syria (2020--2022), where our best estimate of the plausible number of actual fatalities is almost 60\% higher than the actual number reported in the UCDP dataset. For Ethiopia in the same period, the aggregate under-reporting is much smaller, with the mean and median of the distribution of the plausible number of actual fatalities being close to the reported value in the UCDP. However, the model estimates also show that there is an approximate 10\%likelihood that the number of fatalities is under-reported by at least 10,000, and a non-zero likelihood that the number of fatalities is under-reported by at least 50,000. Similarly, there is also an approximate 10\% likelihood that the number of fatalities reported in the UCDP is over-reported by at least 10,000. 

The differences between these two cases illustrate some important aspects of our methodology. According to the UCDP coders, Syria is an example of a country with a good information context; this is reflected by the density of the distribution of the plausible number of fatalities being tightly clustered around the mean and median, with not much variation. In Ethiopia, on the other hand, the distribution is much wider and has a significantly fatter tail. Consistently, insofar as the relative rate of under-reporting is concerned, the UCDP events in Syria (2020--2022) mostly consist of a large number of events with smaller number of fatalities. As our model shows that the relative rate of under-reporting is higher for lower levels of reported fatalities, this also translates into a higher relative rate of under-reporting when aggregating the events. Similarly, a large proportion of fatalities in Ethiopia (75,340) is attributed to a single event, which according to the model suffers from a lower relative rate of under-reporting, but is also less precise. This explains the relatively lower rate of under-reporting in Ethiopia, characterized by fewer but extremely violent events, than in Syria, where there are multiple events associated with a lower number of reported fatalities.

\subsection{Qualitative discussion}

The elicitation approach taught us few relevant lessons on uncertainty in conflict fatalities. The open discussion during the elicitation workshop confirmed the tendency of UCDP to be conservative in their reporting of conflict deaths: in many cases the coders are aware of more deaths associated with a conflict event, but they do not have enough reliable and unbiased evidence to document them according to their coding rules and in line with their definition of conflict. Hence, they generally believe that the actual, unreported number of fatalities is closer to their official `high' estimate than the reported `best' estimate. Especially for cases where there is little available information, the best estimates are almost certain to be severely under-reported.

The workshop also highlighted that the UCDP has a solid understanding of the main drivers of uncertainty at both the sub-national and national level, substantially in line with previous studies. The majority of the team believes that local news tend to be more reliable and accurate than international news and that reports from humanitarian organizations, IGOs or NGOs are more accurate than media sources. New sources of information are regarded as more uncertain than sources with a long history of reporting. Geographic characteristics also affect media bias: urban areas with better access to communication receive better coverage, while reporting for mountainous areas is considerably more uncertain. This may indicate that the positive relationship found in the literature between mountainous terrain and conflict risk may be biased downward (due to the averaging of coefficients between areas with low and high under-reporting), but its statistical significance may also be artificially inflated by the under-reporting bias.

In line with previous studies \citep{Drakos2006JCR}, UCDP coders believe that countries' political regime and freedom of the press largely influence the conflict reporting and hampers media coverage. The violence type (non-state, one-sided, state-based) is associated with different levels of measurement uncertainty, although this association is relative and dependent on the country. As expected, media coverage considerably varies over time, with recent conflicts receiving substantially more attention and coverage than old ones. The type of information provider and their involvement in the conflict has a strong impact on the reliability of the information: each conflict side has incentives to under-report their losses and over-report the deaths among the other party.
The UCDP team is also well aware of the limits of their knowledge: they believe that, especially for bad information contexts and in countries where freedom of the press is severely curtailed, deaths are considerably under-reported, but they cannot have an estimate of the magnitude of this under-reporting, since entire conflict events go undocumented. Although their training and expertise provides them with some intuition of the magnitude of the unreported fatalities, given those that are reported, coders do not have any knowledge of the number of deaths for countries or conflict events for which no reports exist.

\section{Conclusion}


We adopt an elicitation approach to estimate the uncertainty around the number of fatalities reported in the UCDP dataset for state-based, one-sided, and non-state violence, and across different countries and information contexts. The estimated probability distributions show similar patterns across different types of violence, although state-based conflicts appear to be slightly more certain than one-sided and non-state violence for a good information context, and slightly more uncertain for a bad information context. According to our simulations based on UCDP expert assessments, reported fatalities in the UCDP dataset generally suffer from under-reporting, especially at very low and very high levels. Extremely violent cases, with fatalities up to 100,000, are considerably more uncertain than conflict events with fatalities between 20 and 50, and are at risk of both over and under--reporting.

The estimated probability distributions can be used as confidence intervals around the best estimate provided by UCDP, and thus avoid the misuse of low and high estimates as uncertainty boundaries. An improved understanding of the magnitude of measurement uncertainty can foster new efforts to study the broader implications of measurement error for empirical studies of peace and conflict. We know that measurement error induces bias for studies that use news reports for measuring conflict. For example, cellular coverage increases the likelihood that an event is covered by the media, artificially inflating the positive association between the presence of information and communication technologies and the occurrence of violence \citep{Dafoe2015JPR}. Similarly, the relationship between political regime and the incidence of terrorist attacks might be `double-biased' by the reporting error in event datasets: the magnitude of the relationship between regime and terrorism is likely to be biased downward, while the statistical significance of the coefficient is artificially inflated \citep{Drakos2006JCR}. Accounting for measurement uncertainty in conflict data can thus help strengthen the soundness of future empirical studies. Lastly, our estimates of uncertainty can support conflict forecasting exercises that aim at going beyond binary point predictions of conflict occurrence, and rather provide forecasts as probability distribution of conflict fatalities.

This study has some important limitations. First and foremost, our estimates of uncertainty cannot measure how many fatalities are associated with events that are completely missed by reporting.
Post-conflict truth and reconciliations commissions have highlighted that some conflict events are likely to never be publicly recorded, and that only detailed field research could address this gap \citep{Dietrich2020InternationalInteractions}. UCDP experts cannot know exactly how many events do not reach the news, and how many deaths are associated with those events. In other words, the true undocumented number of fatalities is unknowable for large contexts and across countries, and establishing this truth is far beyond the scope of this study. Our goal here is to provide a quantifiable estimate of the approximate distribution of the undocumented deaths. Second, this study does not assess inter-coder reliability: individual variation is a fundamental driver of the responses, and although we control for individual factors, it is possible that part of the uncertainty captured in our estimates is driven by individual propensity to risk, gender, education, experience, and other subjective characteristics of the coders. Further research needs to be done to assess how coding of conflict events vary as a function of individual choices. Lastly, these results cannot be generalized to other conflict event datasets, as the elicitation approach was designed specifically for the UCDP, based on their distinct coding rules and protocols. However, we believe that our approach could easily be replicated for other conflict event datasets, thereby facilitating cross-datasets validation practices, and expanding the generalizability of the results presented here.

\printbibliography

\renewcommand{\thetable}{A-\arabic{table}}
\renewcommand{\thesection}{A-\arabic{section}}
\renewcommand{\thefigure}{A-\arabic{figure}}
\setcounter{table}{0}
\setcounter{section}{0}
\setcounter{figure}{0}

\newpage

\appendix

\section*{Appendix}
\begin{tiny}
\begin{longtable}{p{.04\linewidth}|p{.08\linewidth}|p{.16\linewidth}|p{.12\linewidth}|p{.12\linewidth}|p{.12\linewidth}|p{.04\linewidth}|p{.03\linewidth}}

\textbf{Category } & \textbf{Event type}	&	\textbf{Usage}	&	\textbf{Use of low}	&	\textbf{Use of best}	&	\textbf{Use of high}	&	\textbf{N events} 	&	\textbf{\%}
\\ \hline
1 & Single	&	Only one fatality estimate available from a trusted source	&	Low = best = high	&	Low = best = high	&		Low = best = high	&	224243	&	76.36	\\
1 & Single	&	Conflicting fatalities estimate from trusted sources reporting of a single day event or overnight event	&	Estimate from source that likely under-reports (e.g. government official) &	Lower estimate from credible source	&	Higher estimate from alternative trusted source, or government/rebel claims of fatalities 	&	20746	&	7.06	\\
2 & Single	&	Credible fatality estimates that cannot be firmly linked to the dyad (coding based on UCDP best guess), or estimate from biased but still credible source	&	0	&	Conservative fatality estimate	&	Best fatality estimate	&	5497	&	1.87	\\
2 & Single	&	Credible fatality estimate with probable link to a dyad but slightly uncertain; reasonable claim available only from biased source; second event when there is risk of double counting, with unclear reported date and geography; context specific rules (e.g. snipers killing civilians in Syria coded as high only in state-based since they could be coded as one-sided)	&	0	&	0	&	Fatality estimate reported as high only	&	21058	&	7.17	\\
2 & Single	&	Countries with very limited information; claim from biased source only, but many triangulated sources confirms that the event occurred 	&	0	&	1	&	best fatality estimate reported as high only	&	175	&	0.06	\\
2 & Single	& unclear whether 1 fatality occurred or only injury/capture &	0	&	0	&	1	&	9687	&	3.30	\\
3 & Summary	&	Trusted source with a fatality estimate covering multiple single events, most often within a limited time-period and area, sometimes for the whole dyad-county-year	&	Fatality estimate minus sum of low estimates for associated single events &	Fatality estimate minus sum of best estimates for associated single events	&	Fatality estimate minus the sum of all high estimates in associated single events &	13909	&	4.74	\\
3 & Summary	&	Credible but conflicting estimates of fatalities covering multiple single events, most often within a limited time-period in a limited area &	Reasonable figure by NGO or third party, most often same as above case	&	Reasonable figure by NGO or third party most often same as above case	&	Reasonable higher figure by NGO or third party minus the sum of fatalities in high from single events	&	5542	&	1.89	\\
3 & Summary	&	Summary event for a longer time period and geographical area, e.g. spanning one dyad-county-year	&	0	&	Fatality estimate minus sum of best estimates for associated single events	&	Fatality estimate minus the sum of all high estimates in associated single events	&	502	&	0.17	\\
Summary	&	Countries with very limited reporting; Credible sources confirming at least 25 deaths per year, but no or few associated single events	&	0	&	3 & 25 minus associated fatalities for single events if any &	Most often 0	&	84	&	0.03	\\
3 & Summary	&	Credible information from only one source about event of uncertain type of violence among those coded by UCDP (state-based violence is preferred)	&	0	&	0	&	Best available estimate	&	2458	&	0.84 \\
\hline \\

    \caption{Types of UCDP events and (low, best, high) fatality records. The total number of events is 293,658.}
    \label{tab:UCDP_coding}
\end{longtable}
  \end{tiny}

\newpage

\subsubsection*{Best performing distributions}

\begin{table}[ht]
\centering
\begin{tabular}{rllllllllrr}
\hline
& Distribution & mixture & shifted & ToV & ivs $\theta_1$ & ivs $\theta_2$ & ivs $\theta_3$ & ivs $\theta_4$ & mean score & median score \\
\hline
1 & gumbel & X &  & sb & $\tilde{y}$,$D_i$ & $\tilde{y}$,$D_i$ & $\tilde{y}$,$D_i$,z &  & 0.69 & 0.62 \\
2 & normal & X &  & sb & $\tilde{y}$,$D_i$ & $\tilde{y}$,$D_i$ & $\tilde{y}$,$D_i$ &  & 0.69 & 0.65 \\
3 & poisson & X &  & sb & $\tilde{y}$,$D_i$,z & none & $\tilde{y}$,$D_i$,z &  & 0.78 & 0.72 \\
4 & lognormal & X &  & sb & $\tilde{y}$,z & none & $\tilde{y}$ &  & 0.75 & 0.73 \\
5 & negbin & X &  & sb & $\tilde{y}$ & none & none &  & 0.76 & 0.73 \\
6 & normal & X & X & sb & $\tilde{y}$ & $\tilde{y}$,z & none & $\tilde{y}$,$D_i$ & 1.01 & 0.97 \\
7 & lognormal &  &  & sb & $\tilde{y}$ & none &  &  & 1.06 & 1.01 \\
8 & lognormal & X & X & sb & $\tilde{y}$,z & none & $\tilde{y}$ & $\tilde{y}$ & 1.06 & 1.03 \\
9 & gumbel &  &  & sb & $\tilde{y}$,z & $\tilde{y}$,z &  &  & 1.11 & 1.05 \\
10 & negbin &  &  & sb & $\tilde{y}$,z & none &  &  & 1.10 & 1.10 \\
11 & gumbel & X & X & sb & $\tilde{y}$,$D_i$ & none & $\tilde{y}$ & $\tilde{y}$ & 1.13 & 1.10 \\
12 & normal &  &  & sb & $\tilde{y}$,$D_i$,z & $\tilde{y}$,$D_i$,z &  &  & 1.12 & 1.11 \\
13 & negbin & X & X & sb & $\tilde{y}$ & none & $\tilde{y}$,z & $\tilde{y}$ & 1.10 & 1.12 \\
14 & poisson &  &  & sb & $\tilde{y}$,$D_i$,z & none &  &  & 1.17 & 1.14 \\
15 & poisson & X & X & sb & $\tilde{y}$,z & none & none & $\tilde{y}$ & 1.20 & 1.26 \\
16 & negbin &  & X & sb & $\tilde{y}$,z &  & none & $\tilde{y}$ & 1.51 & 1.73 \\
17 & normal &  & X & sb & $\tilde{y}$,$D_i$ &  & none & $\tilde{y}$ & 1.64 & 1.73 \\
18 & gumbel &  & X & sb & $\tilde{y}$ &  & none & $\tilde{y}$,$D_i$,z & 1.63 & 1.75 \\
19 & lognormal &  & X & sb & $\tilde{y}$,z &  & none & $\tilde{y}$,z & 1.60 & 1.80 \\
20 & poisson &  & X & sb & $\tilde{y}$,$D_i$,z &  & none & $\tilde{y}$,z & 1.74 & 2.00 \\
\hline
\end{tabular}
\caption{LOCO comparison of the best distributions for state-based violence}
\end{table}

\newpage

\begin{table}[ht]
\centering
\begin{tabular}{rllllllllrr}
\hline
& Distribution & mixture & shifted & ToV & ivs $\theta_1$ & ivs $\theta_2$ & ivs $\theta_3$ & ivs $\theta_4$ & mean score & median score \\
\hline
1 & gumbel & X &  & ns & $\tilde{y}$,$D_i$ & $\tilde{y}$,z & $\tilde{y}$,$D_i$ &  & 0.67 & 0.62 \\
2 & normal & X &  & ns & $\tilde{y}$,z & $\tilde{y}$,$D_i$ & $\tilde{y}$,$D_i$ &  & 0.71 & 0.63 \\
3 & lognormal & X &  & ns & $\tilde{y}$ & none & $\tilde{y}$,$D_i$,z &  & 0.72 & 0.67 \\
4 & negbin & X &  & ns & $\tilde{y}$,$D_i$ & none & $D_i$ &  & 0.73 & 0.69 \\
5 & poisson & X &  & ns & $\tilde{y}$,$D_i$ & none & $D_i$,z &  & 0.73 & 0.70 \\
6 & lognormal &  &  & ns & $\tilde{y}$,$D_i$ & $D_i$ &  &  & 1.02 & 1.00 \\
7 & normal & X & X & ns & $\tilde{y}$ & none & $\tilde{y}$,z & $\tilde{y}$ & 1.09 & 1.00 \\
8 & negbin &  &  & ns & $\tilde{y}$ & none &  &  & 1.06 & 1.03 \\
9 & gumbel &  &  & ns & $\tilde{y}$,$D_i$ & $\tilde{y}$ &  &  & 1.07 & 1.04 \\
10 & gumbel & X & X & ns & $\tilde{y}$,$D_i$ & none & $\tilde{y}$,z & $\tilde{y}$ & 1.10 & 1.05 \\
11 & lognormal & X & X & ns & $\tilde{y}$ & none & $\tilde{y}$,z & $\tilde{y}$,z & 1.04 & 1.05 \\
12 & poisson &  &  & ns & $\tilde{y}$,$D_i$,z & none &  &  & 1.10 & 1.09 \\
13 & normal &  &  & ns & $\tilde{y}$,$D_i$,z & $\tilde{y}$ &  &  & 1.12 & 1.10 \\
14 & negbin & X & X & ns & $\tilde{y}$ & none & $\tilde{y}$,z & $\tilde{y}$,z & 1.11 & 1.16 \\
15 & poisson & X & X & ns & $\tilde{y}$,$D_i$,z & none & $\tilde{y}$,z & $\tilde{y}$,$D_i$,z & 1.23 & 1.33 \\
16 & negbin &  & X & ns & $\tilde{y}$,z &  & none & $\tilde{y}$ & 1.51 & 1.70 \\
17 & normal &  & X & ns & $\tilde{y}$,$D_i$ &  & none & $\tilde{y}$ & 1.65 & 1.74 \\
18 & gumbel &  & X & ns & $\tilde{y}$,$D_i$,z &  & none & $\tilde{y}$ & 1.65 & 1.76 \\
19 & lognormal &  & X & ns & $\tilde{y}$,$D_i$ &  & none & $\tilde{y}$ & 1.58 & 1.83 \\
20 & poisson &  & X & ns & $\tilde{y}$,$D_i$ &  & none & $\tilde{y}$,$D_i$,z & 1.76 & 2.00 \\
\hline
\end{tabular}
\caption{LOCO comparison of the best distributions for non-state violence}
\end{table}

\newpage

\begin{table}[ht]
\begin{tabular}{rllllllllrr}
\hline
& Distribution & mixture & shifted & ToV & ivs $\theta_1$ & ivs $\theta_2$ & ivs $\theta_3$ & ivs $\theta_4$ & mean score & median score \\
\hline
1 & gumbel & X &  & os & $\tilde{y}$,$D_i$,z & $\tilde{y}$,z & $\tilde{y}$,$D_i$ &  & 0.62 & 0.57 \\
2 & normal & X &  & os & $\tilde{y}$,$D_i$ & $\tilde{y}$,$D_i$ & $\tilde{y}$,$D_i$,z &  & 0.62 & 0.59 \\
3 & lognormal & X &  & os & $\tilde{y}$,$D_i$ & $D_i$ & $\tilde{y}$,$D_i$,z &  & 0.63 & 0.59 \\
4 & poisson & X &  & os & $\tilde{y}$,$D_i$ & none & $\tilde{y}$,$D_i$,z &  & 0.69 & 0.66 \\
5 & negbin & X &  & os & $\tilde{y}$,$D_i$ & none & $D_i$ &  & 0.70 & 0.69 \\
6 & lognormal &  &  & os & $\tilde{y}$,$D_i$ & $D_i$ &  &  & 0.95 & 0.92 \\
7 & gumbel &  &  & os & $\tilde{y}$,$D_i$ & $\tilde{y}$,z &  &  & 1.02 & 0.98 \\
8 & normal & X & X & os & $\tilde{y}$,z & none & $\tilde{y}$,z & $\tilde{y}$ & 1.07 & 1.01 \\
9 & negbin &  &  & os & $\tilde{y}$ & none &  &  & 1.04 & 1.02 \\
10 & normal &  &  & os & $\tilde{y}$,$D_i$ & $\tilde{y}$ &  &  & 1.09 & 1.07 \\
11 & lognormal & X & X & os & $\tilde{y}$,z & none & $\tilde{y}$,z & $\tilde{y}$,z & 1.05 & 1.07 \\
12 & poisson &  &  & os & $\tilde{y}$,$D_i$ & none &  &  & 1.08 & 1.09 \\
13 & gumbel & X & X & os & $\tilde{y}$,$D_i$,z & none & $\tilde{y}$ & $\tilde{y}$ & 1.11 & 1.10 \\
14 & negbin & X & X & os & $\tilde{y}$ & none & $\tilde{y}$,z & $\tilde{y}$,$D_i$ & 1.12 & 1.14 \\
15 & poisson & X & X & os & $\tilde{y}$ & none & $\tilde{y}$,z & $\tilde{y}$,$D_i$ & 1.24 & 1.29 \\
16 & negbin &  & X & os & $\tilde{y}$,z &  & none & $\tilde{y}$ & 1.50 & 1.71 \\
17 & normal &  & X & os & $\tilde{y}$,$D_i$,z &  & none & $\tilde{y}$ & 1.67 & 1.81 \\
18 & lognormal &  & X & os & $\tilde{y}$,$D_i$ & $\tilde{y}$ & none & $\tilde{y}$ & 1.58 & 1.83 \\
19 & gumbel &  & X & os & $\tilde{y}$,z &  & none & $\tilde{y}$,$D_i$ & 1.65 & 1.87 \\
20 & poisson &  & X & os & $\tilde{y}$,$D_i$,z &  & none & $\tilde{y}$ & 1.72 & 2.00 \\
\hline
\end{tabular}
\caption{LOCO comparison of the best distributions for one-sided violence}
\end{table}

\newpage

\begin{table}[ht]
\centering
\begin{tabular}{rllllllllrr}
\hline
& Distribution & mixture & shifted & ToV & ivs $\theta_1$ & ivs $\theta_2$ & ivs $\theta_3$ & ivs $\theta_4$ & mean score & median score \\
\hline
1 & gumbel & X &  & all & $\tilde{y}$,$D_i$ & $\tilde{y}$ & $\tilde{y}$,$D_i$ &  & 0.66 & 0.60 \\
2 & normal & X &  & all & $\tilde{y}$,$D_i$ & $\tilde{y}$,$D_i$ & $\tilde{y}$,$D_i$ &  & 0.65 & 0.63 \\
3 & lognormal & X &  & all & $\tilde{y}$ & none & $D_i$ &  & 0.70 & 0.67 \\
4 & poisson & X &  & all & $\tilde{y}$,$D_i$ & none & none &  & 0.73 & 0.69 \\
5 & negbin & X &  & all & $\tilde{y}$,$D_i$ & none & $D_i$ &  & 0.73 & 0.71 \\
6 & lognormal &  &  & all & $\tilde{y}$ & none &  &  & 1.02 & 0.99 \\
7 & normal & X & X & all & $\tilde{y}$ & none & $\tilde{y}$,z & $\tilde{y}$ & 1.08 & 1.01 \\
8 & gumbel &  &  & all & $\tilde{y}$,$D_i$,z & $\tilde{y}$,z &  &  & 1.06 & 1.02 \\
9 & negbin &  &  & all & $\tilde{y}$ & none &  &  & 1.06 & 1.06 \\
10 & lognormal & X & X & all & $\tilde{y}$ & $\tilde{y}$,$D_i$ & $\tilde{y}$,z & $\tilde{y}$ & 1.03 & 1.06 \\
11 & gumbel & X & X & all & $\tilde{y}$,$D_i$,z & none & $\tilde{y}$ & $\tilde{y}$,z & 1.11 & 1.08 \\
12 & normal &  &  & all & $\tilde{y}$,$D_i$ & $\tilde{y}$,z &  &  & 1.12 & 1.10 \\
13 & poisson &  &  & all & $\tilde{y}$,$D_i$,z & none &  &  & 1.12 & 1.11 \\
14 & negbin & X & X & all & $\tilde{y}$ & none & $\tilde{y}$,z & $\tilde{y}$,$D_i$,z & 1.12 & 1.14 \\
15 & poisson & X & X & all & $\tilde{y}$ & none & $\tilde{y}$,z & $\tilde{y}$ & 1.21 & 1.29 \\
16 & negbin &  & X & all & $\tilde{y}$ &  & none & $\tilde{y}$ & 1.50 & 1.71 \\
17 & normal &  & X & all & $\tilde{y}$,$D_i$,z &  & none & $\tilde{y}$,z & 1.65 & 1.74 \\
18 & gumbel &  & X & all & $\tilde{y}$,$D_i$ &  & none & $\tilde{y}$ & 1.67 & 1.79 \\
19 & lognormal &  & X & all & $\tilde{y}$,$D_i$ &  & none & $\tilde{y}$ & 1.59 & 1.82 \\
20 & poisson &  & X & all & $\tilde{y}$,$D_i$ &  & none & $\tilde{y}$,z & 1.74 & 2.00 \\
\hline
\end{tabular}
\caption{LOCO comparison of the best distributions for all types of violence}
\end{table}

\newpage

\subsection*{Model selection}

\begin{table}[ht]
\centering
\begin{tabular}{rllllrrr}
  \hline
& ToV & ivs $\theta_1$ & ivs $\theta_2$ & ivs $\theta_3$ & mean score & median score & Rel. increase median \\
  \hline
1 & sb & $\tilde{y}$,$D_i$ & $\tilde{y}$,$D_i$ & $\tilde{y}$,$D_i$,z & 0.69 & 0.62 & 1.00 \\
  2 & sb & $\tilde{y}$,$D_i$ & $\tilde{y}$,$D_i$,z & $\tilde{y}$,$D_i$,z & 0.69 & 0.63 & 1.01 \\
  3 & sb & $\tilde{y}$,$D_i$ & $\tilde{y}$,$D_i$ & $\tilde{y}$,z & 0.69 & 0.63 & 1.01 \\
  4 & sb & $\tilde{y}$,$D_i$ & $\tilde{y}$,$D_i$,z & $\tilde{y}$,z & 0.69 & 0.63 & 1.01 \\
  \bf{5} & \bf{sb} & \bf{$\tilde{y}$,$D_i$} & \bf{$\tilde{y}$,$D_i$} & \bf{$\tilde{y}$,$D_i$} & \bf{0.68} & \bf{0.63} & \bf{1.01} \\
  6 & sb & $\tilde{y}$,$D_i$ & $\tilde{y}$,$D_i$,z & $\tilde{y}$,$D_i$ & 0.68 & 0.63 & 1.02 \\
  7 & sb & $\tilde{y}$,$D_i$ & $\tilde{y}$,$D_i$ & $D_i$ & 0.69 & 0.63 & 1.02 \\
  8 & sb & $\tilde{y}$,$D_i$ & $\tilde{y}$ & $\tilde{y}$,$D_i$,z & 0.70 & 0.63 & 1.02 \\
  9 & sb & $\tilde{y}$,$D_i$ & $\tilde{y}$,$D_i$,z & $\tilde{y}$ & 0.69 & 0.63 & 1.02 \\
  10 & sb & $\tilde{y}$,$D_i$,z & $\tilde{y}$,$D_i$,z & $\tilde{y}$,$D_i$,z & 0.69 & 0.64 & 1.02 \\
  11 & sb & $\tilde{y}$,$D_i$ & $\tilde{y}$ & $\tilde{y}$,z & 0.70 & 0.64 & 1.02 \\
  12 & sb & $\tilde{y}$,$D_i$ & $\tilde{y}$,$D_i$,z & none & 0.69 & 0.64 & 1.02 \\
  13 & sb & $\tilde{y}$,$D_i$ & $\tilde{y}$,z & $\tilde{y}$,$D_i$,z & 0.69 & 0.64 & 1.02 \\
  14 & sb & $\tilde{y}$,$D_i$ & $\tilde{y}$,$D_i$,z & $D_i$ & 0.69 & 0.64 & 1.02 \\
  15 & sb & $\tilde{y}$,$D_i$ & $\tilde{y}$,$D_i$ & $\tilde{y}$ & 0.69 & 0.64 & 1.02 \\
  16 & sb & $\tilde{y}$,$D_i$ & $\tilde{y}$ & $\tilde{y}$,$D_i$ & 0.69 & 0.64 & 1.02 \\
  17 & sb & $\tilde{y}$,$D_i$ & $\tilde{y}$ & $\tilde{y}$ & 0.69 & 0.64 & 1.02 \\
  18 & sb & $\tilde{y}$,$D_i$ & $\tilde{y}$,z & $\tilde{y}$,z & 0.70 & 0.64 & 1.02 \\
  19 & sb & $\tilde{y}$,$D_i$ & $\tilde{y}$,$D_i$,z & $D_i$,z & 0.69 & 0.64 & 1.02 \\
  20 & sb & $\tilde{y}$,$D_i$,z & $\tilde{y}$,$D_i$ & $\tilde{y}$,$D_i$,z & 0.69 & 0.64 & 1.02 \\
  21 & sb & $\tilde{y}$,$D_i$,z & $\tilde{y}$,$D_i$,z & $\tilde{y}$,z & 0.69 & 0.64 & 1.02 \\
  22 & sb & $\tilde{y}$,$D_i$ & $\tilde{y}$ & $D_i$ & 0.69 & 0.64 & 1.02 \\
  23 & sb & $\tilde{y}$,$D_i$ & $\tilde{y}$ & $D_i$,z & 0.70 & 0.64 & 1.02 \\
  24 & sb & $\tilde{y}$,$D_i$ & $\tilde{y}$,$D_i$ & $D_i$,z & 0.69 & 0.64 & 1.03 \\
  25 & sb & $\tilde{y}$,$D_i$,z & $\tilde{y}$,$D_i$,z & $\tilde{y}$,$D_i$ & 0.69 & 0.64 & 1.03 \\
   \hline
\end{tabular}
\caption{LOCO comparison of the gumbel mixture models for state-based violence}
\end{table}

\newpage

\begin{table}[ht]
\centering
\begin{tabular}{rllllrrr}
  \hline
& ToV & ivs $\theta_1$ & ivs $\theta_2$ & ivs $\theta_3$ & mean score & median score & Rel. increase median \\
  \hline
1 & ns & $\tilde{y}$,$D_i$ & $\tilde{y}$,z & $\tilde{y}$,$D_i$ & 0.67 & 0.62 & 1.00 \\
  2 & ns & $\tilde{y}$,$D_i$ & $\tilde{y}$ & $\tilde{y}$,$D_i$ & 0.67 & 0.62 & 1.00 \\
  3 & ns & $\tilde{y}$,$D_i$ & $\tilde{y}$ & $\tilde{y}$,$D_i$,z & 0.67 & 0.62 & 1.00 \\
  4 & ns & $\tilde{y}$,$D_i$ & $\tilde{y}$,$D_i$,z & $\tilde{y}$,$D_i$ & 0.67 & 0.62 & 1.00 \\
  5 & ns & $\tilde{y}$,$D_i$,z & $\tilde{y}$,z & $\tilde{y}$,$D_i$ & 0.67 & 0.62 & 1.00 \\
  6 & ns & $\tilde{y}$,$D_i$ & $\tilde{y}$ & $D_i$ & 0.67 & 0.62 & 1.00 \\
  7 & ns & $\tilde{y}$,$D_i$ & $\tilde{y}$ & $D_i$,z & 0.67 & 0.62 & 1.01 \\
  8 & ns & $\tilde{y}$,$D_i$ & $\tilde{y}$,z & $\tilde{y}$,$D_i$,z & 0.67 & 0.62 & 1.01 \\
  9 & ns & $\tilde{y}$,$D_i$,z & $\tilde{y}$ & $\tilde{y}$,$D_i$ & 0.67 & 0.62 & 1.01 \\
  10 & ns & $\tilde{y}$,$D_i$ & $\tilde{y}$,z & $D_i$ & 0.67 & 0.62 & 1.01 \\
  11 & ns & $\tilde{y}$,$D_i$ & $\tilde{y}$,z & $D_i$,z & 0.67 & 0.62 & 1.01 \\
  12 & ns & $\tilde{y}$,$D_i$,z & $\tilde{y}$ & $D_i$,z & 0.67 & 0.62 & 1.01 \\
  \bf{13} & \bf{os} & \bf{$\tilde{y}$,$D_i$} & \bf{$\tilde{y}$,$D_i$} & \bf{$\tilde{y}$,$D_i$} & \bf{0.67} & \bf{0.62} & \bf{1.01} \\
  14 & ns & $\tilde{y}$,$D_i$,z & $\tilde{y}$ & $\tilde{y}$,$D_i$,z & 0.67 & 0.62 & 1.01 \\
  15 & ns & $\tilde{y}$,$D_i$,z & $\tilde{y}$,z & $D_i$ & 0.67 & 0.62 & 1.01 \\
  16 & ns & $\tilde{y}$,$D_i$ & $\tilde{y}$ & $\tilde{y}$,z & 0.68 & 0.62 & 1.01 \\
  17 & ns & $\tilde{y}$,$D_i$,z & $\tilde{y}$ & $D_i$ & 0.67 & 0.63 & 1.01 \\
  18 & ns & $\tilde{y}$,$D_i$,z & $\tilde{y}$,$D_i$ & $\tilde{y}$,$D_i$ & 0.67 & 0.63 & 1.01 \\
  19 & ns & $\tilde{y}$,$D_i$,z & $\tilde{y}$,z & $D_i$,z & 0.67 & 0.63 & 1.01 \\
  20 & ns & $\tilde{y}$,$D_i$,z & $\tilde{y}$,$D_i$,z & $\tilde{y}$,$D_i$ & 0.67 & 0.63 & 1.01 \\
  21 & ns & $\tilde{y}$,$D_i$ & $\tilde{y}$,$D_i$ & $\tilde{y}$,$D_i$,z & 0.67 & 0.63 & 1.01 \\
  22 & ns & $\tilde{y}$,$D_i$ & $\tilde{y}$,z & $\tilde{y}$,z & 0.68 & 0.63 & 1.01 \\
  23 & ns & $\tilde{y}$,$D_i$ & $\tilde{y}$ & none & 0.68 & 0.63 & 1.01 \\
  24 & ns & $\tilde{y}$,$D_i$ & $\tilde{y}$,$D_i$,z & $D_i$ & 0.67 & 0.63 & 1.01 \\
  25 & ns & $\tilde{y}$,$D_i$ & $\tilde{y}$,$D_i$ & $D_i$ & 0.67 & 0.63 & 1.02 \\
   \hline
\end{tabular}
\caption{LOCO comparison of the gumbel mixture models for non-state violence}
\end{table}

\newpage

\begin{table}
\begin{tabular}{rllllrrr}
  \hline
& ToV & ivs $\theta_1$ & ivs $\theta_2$ & ivs $\theta_3$ & mean score & median score & Rel. increase median \\
  \hline
1 & os & $\tilde{y}$,$D_i$,z & $\tilde{y}$,z & $\tilde{y}$,$D_i$ & 0.62 & 0.57 & 1.00 \\
  2 & os & $\tilde{y}$,$D_i$ & $\tilde{y}$ & $D_i$ & 0.61 & 0.57 & 1.00 \\
  3 & os & $\tilde{y}$,$D_i$,z & $\tilde{y}$ & $\tilde{y}$,$D_i$ & 0.61 & 0.57 & 1.00 \\
  4 & os & $\tilde{y}$,$D_i$,z & $\tilde{y}$,z & $D_i$ & 0.62 & 0.57 & 1.00 \\
  5 & os & $\tilde{y}$,$D_i$ & $\tilde{y}$,z & $D_i$ & 0.62 & 0.57 & 1.01 \\
  6 & os & $\tilde{y}$,$D_i$ & $\tilde{y}$ & none & 0.63 & 0.57 & 1.01 \\
  7 & os & $\tilde{y}$,$D_i$ & $\tilde{y}$ & $\tilde{y}$,$D_i$ & 0.61 & 0.57 & 1.01 \\
  8 & os & $\tilde{y}$,$D_i$,z & $\tilde{y}$ & $D_i$ & 0.62 & 0.57 & 1.01 \\
  9 & os & $\tilde{y}$,$D_i$ & $\tilde{y}$,z & none & 0.63 & 0.57 & 1.01 \\
  10 & os & $\tilde{y}$,$D_i$ & $\tilde{y}$,z & $\tilde{y}$,$D_i$ & 0.62 & 0.57 & 1.01 \\
  11 & os & $\tilde{y}$,$D_i$,z & $\tilde{y}$ & none & 0.63 & 0.57 & 1.01 \\
  12 & os & $\tilde{y}$,$D_i$,z & $\tilde{y}$ & $\tilde{y}$ & 0.62 & 0.57 & 1.01 \\
  13 & os & $\tilde{y}$,$D_i$,z & $\tilde{y}$,z & $\tilde{y}$ & 0.63 & 0.57 & 1.01 \\
  14 & os & $\tilde{y}$,$D_i$,z & $\tilde{y}$,z & none & 0.63 & 0.57 & 1.01 \\
  15 & os & $\tilde{y}$,$D_i$ & $\tilde{y}$ & $\tilde{y}$ & 0.62 & 0.57 & 1.01 \\
  16 & os & $\tilde{y}$,$D_i$ & $\tilde{y}$,z & $\tilde{y}$ & 0.62 & 0.58 & 1.02 \\
  17 & os & $\tilde{y}$,$D_i$,z & $\tilde{y}$ & $D_i$,z & 0.62 & 0.58 & 1.02 \\
  18 & os & $\tilde{y}$,$D_i$ & $\tilde{y}$ & $D_i$,z & 0.61 & 0.58 & 1.02 \\
  19 & os & $\tilde{y}$,$D_i$,z & $\tilde{y}$,z & $D_i$,z & 0.62 & 0.58 & 1.02 \\
  20 & os & $\tilde{y}$,$D_i$ & $\tilde{y}$,z & $D_i$,z & 0.62 & 0.58 & 1.02 \\
  21 & os & $\tilde{y}$,$D_i$ & $\tilde{y}$ & $\tilde{y}$,z & 0.62 & 0.58 & 1.02 \\
  \bf{22} & \bf{os} & \bf{$\tilde{y}$,$D_i$} & \bf{$\tilde{y}$,$D_i$} & \bf{$\tilde{y}$,$D_i$} & \bf{0.62} & \bf{0.58} & \bf{1.02} \\
  23 & os & $\tilde{y}$,$D_i$,z & $\tilde{y}$,$D_i$ & $\tilde{y}$,$D_i$ & 0.62 & 0.58 & 1.02 \\
  24 & os & $\tilde{y}$,$D_i$,z & $\tilde{y}$,$D_i$ & $\tilde{y}$,$D_i$,z & 0.62 & 0.58 & 1.02 \\
  25 & os & $\tilde{y}$,$D_i$,z & $\tilde{y}$,$D_i$,z & $\tilde{y}$,$D_i$ & 0.62 & 0.58 & 1.02 \\
   \hline
\end{tabular}
\caption{LOCO comparison of the gumbel mixture models for one-sided violence}
\end{table}

\newpage

\begin{table}[ht]
\centering
\begin{tabular}{rllllrrr}
  \hline
& ToV & ivs $\theta_1$ & ivs $\theta_2$ & ivs $\theta_3$ & mean score & median score & Rel. increase median \\
  \hline
1 & all & $\tilde{y}$,$D_i$ & $\tilde{y}$ & $\tilde{y}$,$D_i$ & 0.66 & 0.60 & 1.00 \\
  2 & all & $\tilde{y}$,$D_i$ & $\tilde{y}$,z & $\tilde{y}$,$D_i$ & 0.66 & 0.60 & 1.00 \\
  3 & all & $\tilde{y}$,$D_i$ & $\tilde{y}$,z & $D_i$ & 0.66 & 0.60 & 1.00 \\
  4 & all & $\tilde{y}$,$D_i$,z & $\tilde{y}$ & $\tilde{y}$,$D_i$ & 0.66 & 0.60 & 1.01 \\
  5 & all & $\tilde{y}$,$D_i$,z & $\tilde{y}$,z & $\tilde{y}$,$D_i$ & 0.66 & 0.60 & 1.01 \\
  6 & all & $\tilde{y}$,$D_i$ & $\tilde{y}$ & $D_i$ & 0.66 & 0.60 & 1.01 \\
  7 & all & $\tilde{y}$,$D_i$,z & $\tilde{y}$ & $D_i$ & 0.66 & 0.61 & 1.01 \\
  8 & all & $\tilde{y}$,$D_i$,z & $\tilde{y}$,z & $D_i$ & 0.66 & 0.61 & 1.01 \\
  9 & all & $\tilde{y}$,$D_i$ & $\tilde{y}$ & $\tilde{y}$,$D_i$,z & 0.66 & 0.61 & 1.01 \\
  10 & all & $\tilde{y}$,$D_i$ & $\tilde{y}$ & $\tilde{y}$ & 0.66 & 0.61 & 1.01 \\
 \bf{11} & \bf{sb} & \bf{$\tilde{y}$,$D_i$} & \bf{$\tilde{y}$,$D_i$} & \bf{$\tilde{y}$,$D_i$} & \bf{0.65} & \bf{0.61} & \bf{1.01} \\
  12 & all & $\tilde{y}$,$D_i$ & $\tilde{y}$,z & $\tilde{y}$ & 0.66 & 0.61 & 1.01 \\
  13 & all & $\tilde{y}$,$D_i$,z & $\tilde{y}$,$D_i$ & $\tilde{y}$,$D_i$,z & 0.66 & 0.61 & 1.01 \\
  14 & all & $\tilde{y}$,$D_i$ & $\tilde{y}$,z & $\tilde{y}$,$D_i$,z & 0.66 & 0.61 & 1.01 \\
  15 & all & $\tilde{y}$,$D_i$,z & $\tilde{y}$,z & $\tilde{y}$ & 0.67 & 0.61 & 1.01 \\
  16 & all & $\tilde{y}$,$D_i$,z & $\tilde{y}$ & none & 0.67 & 0.61 & 1.02 \\
  17 & all & $\tilde{y}$,$D_i$ & $\tilde{y}$ & $\tilde{y}$,z & 0.67 & 0.61 & 1.02 \\
  18 & all & $\tilde{y}$,$D_i$ & $\tilde{y}$,z & $\tilde{y}$,z & 0.67 & 0.61 & 1.02 \\
  19 & all & $\tilde{y}$,$D_i$,z & $\tilde{y}$,z & none & 0.67 & 0.61 & 1.02 \\
  20 & all & $\tilde{y}$,$D_i$,z & $\tilde{y}$,$D_i$ & $\tilde{y}$,$D_i$ & 0.65 & 0.61 & 1.02 \\
  21 & all & $\tilde{y}$,$D_i$ & $\tilde{y}$,$D_i$ & $\tilde{y}$,$D_i$,z & 0.66 & 0.61 & 1.02 \\
  22 & all & $\tilde{y}$,$D_i$ & $\tilde{y}$ & $D_i$,z & 0.66 & 0.61 & 1.02 \\
  23 & all & $\tilde{y}$,$D_i$,z & $\tilde{y}$ & $\tilde{y}$,$D_i$,z & 0.66 & 0.61 & 1.02 \\
  24 & all & $\tilde{y}$,$D_i$,z & $\tilde{y}$,$D_i$,z & $\tilde{y}$,$D_i$,z & 0.66 & 0.61 & 1.02 \\
  25 & all & $\tilde{y}$,$D_i$ & $\tilde{y}$,$D_i$ & $D_i$,z & 0.66 & 0.61 & 1.02 \\
   \hline
\end{tabular}
\caption{LOCO comparison of the gumbel mixture models for all types of violence}
\end{table}

\newpage





\end{document}